\documentclass[twocolumn,preprintnumbers,aps]{revtex4-1}
\usepackage{amsmath}
\usepackage{graphicx}
\usepackage{color}
\usepackage{subfigure}
\usepackage{hyperref}

\usepackage{siunitx}
\DeclareSIUnit{\atmosphere}{atm}

\newcommand{\BAN}{\ensuremath{B_{1g}\,}}
\newcommand{\BN}{\ensuremath{B_{2g}\,}}

\newcommand{\Ts}{\ensuremath{T^{\ast}\,}}
\newcommand{\Tc}{\ensuremath{T_{\rm c}\,}}
\newcommand{\cm}{\ensuremath{{\rm cm}^{-1}}}

\graphicspath{{images/}{../images/}}

\begin{document}

\title{Exploration of the Hg-based cuprate superconductors by Raman spectroscopy under hydrostatic pressure}
\author{N. Auvray$^1$, B. Loret $^1$, S. Chibani $^1$, R. Grasset $^2$, Y. Guarnelli $^3$, P. Parisiades$^3$, A. Forget$^4$, D. Colson$^4$, M. Cazayous $^1$,Y. Gallais $^1$, A. Sacuto$^*$}
\affiliation{Universit\'e de Paris, Laboratoire Mat\'eriaux et Ph\'enom$\grave{e}$nes Quantiques, CNRS (UMR 7162), 75013 Paris, France\\
$^2$ Laboratoire des Solides Irradi\'es, Institut Rayonnement Mati$\grave{e}$re de Saclay (IRAMIS), CEA, Ecole Polytechnique, 91128 PALAISEAU Cedex, France\\
$^3$ Sorbonne Universit\'e, Institut de min\'eralogie, de physique des mat\'eriaux et de cosmochimie, CNRS/MNHN/IRD (UMR 7590), 75005 Paris, France\\
$^4$ Universit\'e Paris-Saclay, CEA, CNRS, SPEC, 91191, Gif-sur-Yvette, France\\
}

\date{\today}

\begin{abstract}
The superconducting phase of the $\mathrm{HgBa}_2\mathrm{CuO}_{4+\delta}$ (Hg-1201) and $\mathrm{HgBa}_2\mathrm{Ca}_2\mathrm{Cu}_3\mathrm{O}_{8+\delta}$ (Hg-1223) cuprates has been investigated by Raman spectroscopy under hydrostatic pressure. Our analysis reveals that the increase of \Tc with pressure is slower in Hg-1223 cuprate compared to the Hg-1201 due to a charge carrier concentration imbalance (accentuated by pressure) between the  $\mathrm{CuO}_2$ layers of Hg-1223. We find that the energy variation under pressure of the apical oxygen mode from which the charge carriers are transferred to the $\mathrm{CuO}_2$ layers, is the same for both the Hg-1223 and Hg-1223 cuprates and it is controlled by the inter-layer compressibility. At last, we show that the binding energy of the Cooper pairs related to the maximum amplitude of the $d-$ wave superconducting gap at the anti-nodes, does not follow \Tc with pressure. It decreases while \Tc increases. In the particular case of Hg-1201, the binding energy collapses from 10 to 2 $K_B\Tc$ as the pressure increases up to 10 GPa. These direct spectroscopic observations joined to the fact that the binding energy of the Cooper pairs at the anti-nodes does not follow \Tc either with doping, raises the question of its link with the pseudogap energy scale which follows the same trend with doping.

\end{abstract}

\maketitle

\section{I. Introduction}

High-\Tc cuprate superconductors are one of the iconic quantum materials \cite{Norman2011,Keimer2015}. Although discovered more than 35 years ago, the complexity of their physics remains misunderstood. It calls for new concepts where the orders of matter are no longer independent of each other as in traditional materials but they are intertwined \cite{Fradkin2014}. In order to understand their physics, many studies have already been carried out as a function of temperature $T$ and carriers concentration via the hole doping, $p$, leading to their ($T-p$) phase diagram \cite{Lee06,Keimer2015}. It presents an insulating anti-ferromagnetic phase at low doping.  As the doping increases, an intermediate phase between the insulator and the metal called the pseudogap phase emerges below \Ts which harbors many orders of matter that appear to be interconnected. Some of them break translational invariance (charge density wave order), others time reversal invariance (current loops order) or $C_4$ rotational invariance (nematic order)\cite{Hanaguri04,Wu11,Fujita14,Comin2016,Arpaia2019,Fauque06,Daou2010,Proust2019,Sato2017,Auvray2019}. At lower temperature below a critical temperature \Tc, the superconducting phase arises. \Tc exhibits a dome like shape. The top of the dome called the optimal doping, separates the under-doped from the over-doped regime. The physics behind this phase diagram remains widely debated and calls for the development of new experiments and theoretical investigations \cite{Lee06,Scalapino2012,Sachdev13,Fradkin2014,Wang2015,Caprara2017,Wu2018,Chakraborty2019,Choubey2020}. 
\par
\begin{figure}[ht]
\includegraphics[scale=0.18]{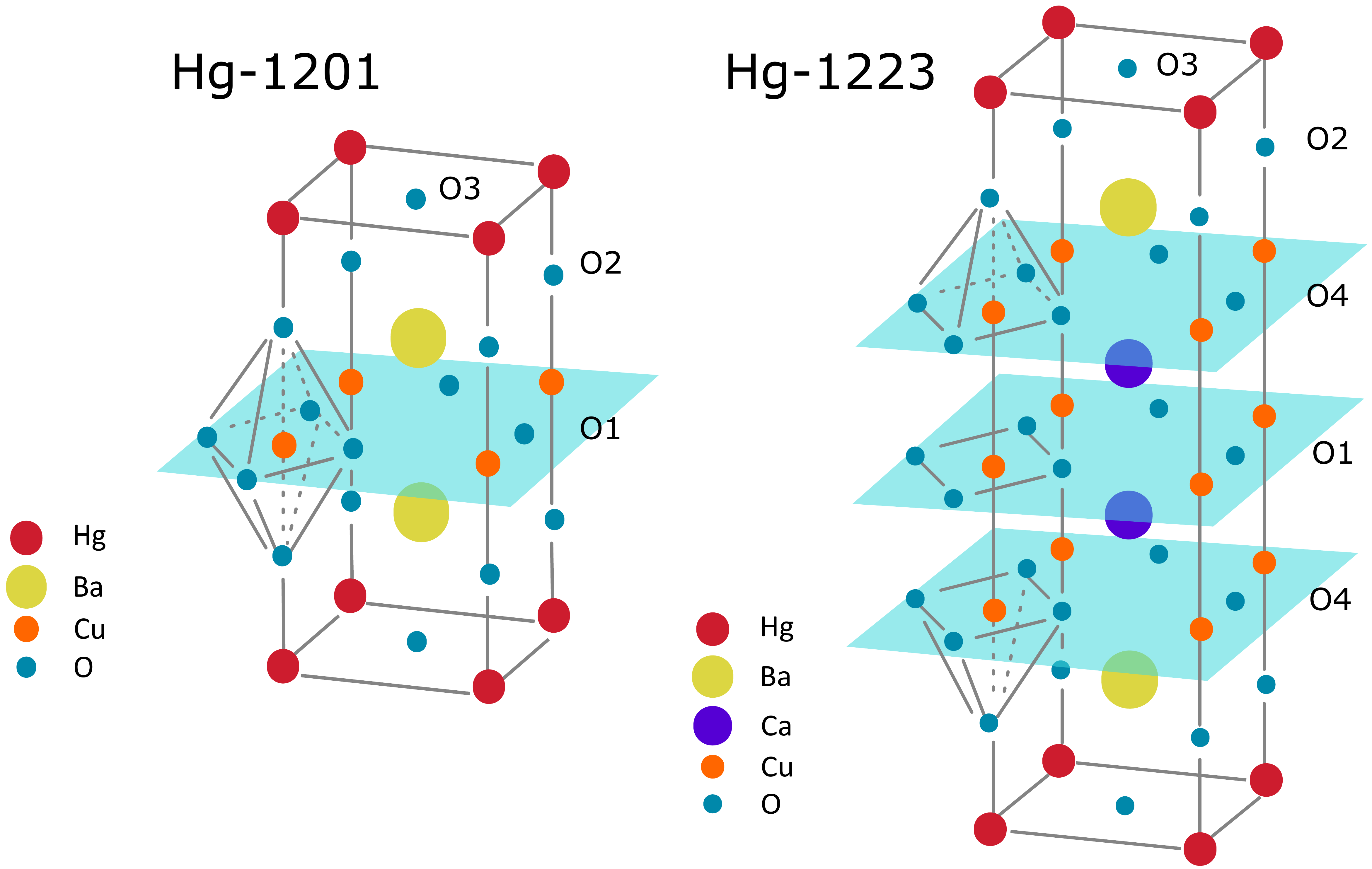}
\caption{Schemtic representation of the tetragonal crystal structures (a) Hg-1201 and (b) Hg-1223. the oxygen atoms O1 and O4 are respectively related to the inner and outer $\mathrm{CuO}_2$ planes, O2 are the apical oxygen atoms, O3 are the oxygen atoms in excess inserted by the chemical doping in the $\mathrm{HgO}$ plane.}
\label{fig1}
\end{figure}

In order to get a better understanding of the cuprates physics and in particular their superconducting phase, we carried out Raman scattering measurements under hydrostatic pressure at low temperature on the Hg-based cuprates which have the most spectacular variation of \Tc with pressure. They exhibit an increase in \Tc of more than \num{25} K for a pressure of \SI{25}{\giga\pascal}, i.e. on average, an increase of \num{1} K per GPa \cite{Chu1993,nunez1993,Gao1994,Antipov2002}. We will focus on the $\mathrm{HgBa}_2\mathrm{CuO}_{4+\delta}$ (Hg-1201) and  $\mathrm{HgBa}_2\mathrm{Ca}_2\mathrm{Cu}_3\mathrm{O}_{8+\delta}$ (Hg-1223) compounds which have respectively a maximum \Tc $\approx$ 95 K and $\approx$ 135 K at ambient pressure ($\approx 0$ GPa) \cite{Putilin1993,Schilling1993}. The structural unit cells are respectively made of one single and three contiguous $\mathrm{CuO}_2$ planes, surrounded by blocks made up of  $\mathrm{HgO}$ and $\mathrm{BaO}$ layers (cf. Fig.~\ref{fig1}). These blocks are called charge reservoirs because the introduction of oxygen atoms within the $\mathrm{HgO}$ plane generates a charge transfer via the $\mathrm{BaO}$ plane towards the $\mathrm{CuO}_2$. This oxygen doping introduces hole charge carriers in the $\mathrm{CuO}_2$ plane \cite{Antipov2002,Fukuoka1997,Kotegawa2002,Mukuda2012}.

%%%%%%%%%%%%%%%%%%%%%%%%%%%%%%%%%%%%%%%%%%%%%%%%%%%%%%%%%%%%%%%%%%%%%%%%%%%%%%%%%%%%%%%%%%%%%%%%%%%%%%%%%%%%%%%%
%%%%%%%%%%%%%%%%%%%%%%%%%%%%%%%%%%%%%%%%%%%%%%%%%%%%%%%%%%%%%%%%%%%%%%%%%%%%%%%%%%%%%%%%%%%%%%%%%%%%%%%%%%%%%%%%

In the cuprates Raman scattering has been used extensively to track the energy scales of the phonons, the magnetic excitations, the superconducting gap, the pseudogap \cite{Devereaux2007,Blumberg1997,Opel2000,LeTacon2006,Blanc2010,li2012,benhabib15,Loret2017a,Loret2018} or more recently the charge density wave gap \cite{Loret2019,Li2019,Loret2020,Wang2020}. Since Raman is a two photon scattering process, by controlling the incoming and outgoing photon polarizations, one can selectively probe both the lattice and the electronic excitations in different symmetries. In the superconducting phase, we will be focusing on the relationships that can be unveiled with pressure between lattice dynamics and \Tc and also between the binding energy of the Cooper pairs associated with the $d$-wave superconducting (SC) gap and \Tc. We will show that the increase of \Tc with pressure is reduced in Hg-1223 below 10 GPa in comparison to the Hg-1201 compound due to an increase of the charge carriers concentration imbalance between the inner and outer $\mathrm{CuO}_2$ planes of Hg-1223 with pressure. We find that the evolution under pressure of frequency of the the apical oxygen, by which the charge transfer takes place, is mainly controlled by the inter-layer contraction. At last, we show that the binding energy of the Cooper pairs related to the maximum amplitude of the $d-$ wave SC gap along the principal axis of the Brillouin zone (BZ) called the anti-nodal region, does not follow \Tc with pressure. It decrease while \Tc increases with pressure. Our findings, together with previous investigations that showed the binding energy of the Cooper pairs at the anti-nodes decreases as \Tc increases with doping and follows the same trend as the pseudogap energy scale \cite{Tallon01,LeTacon2006,Kanigel2006,Fischer07,Bernhard2008,Munnikes2011,Loret2020}, raise the question of its link to the pseudogap energy scale.

\section{II. Experimental Methods}

\subsection{A. Crystal growth and characterization}

\begin{figure}[ht!]
\includegraphics[scale=0.32]{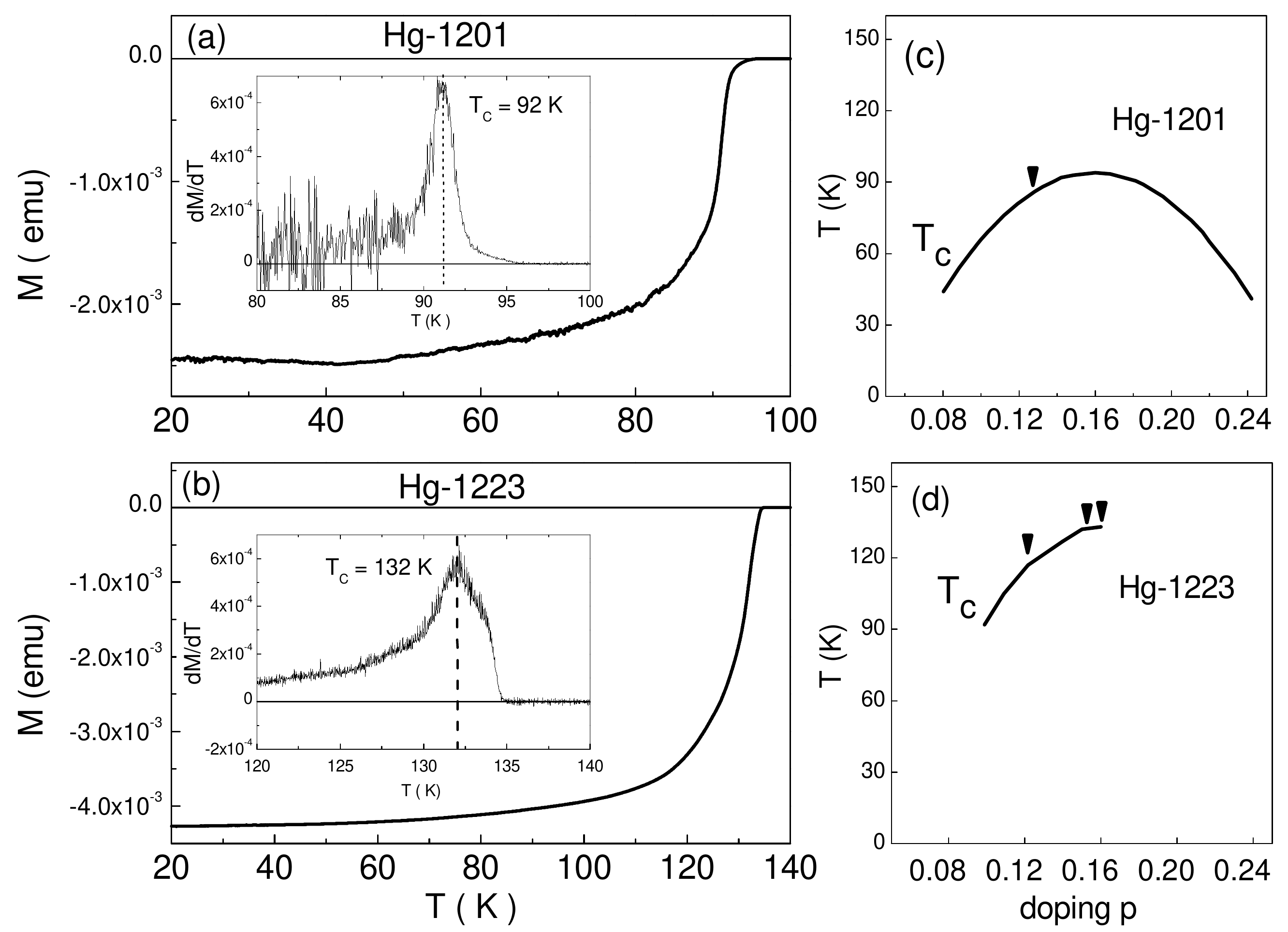}
\caption{DC magnetization curves as a function of temperature of two sets of (a) Hg-1201 and (b) Hg-1223 single crystals. These crystals have been used for Raman measurements under pressureThe first derivatives of the magnetization curves are reported in the insets to underline the \Tc values and their full width at half maximum. The \Tc values of the mercurate single crystals measured by Raman spectroscopy are reported in (c) for the Hg-1201 and (d) Hg-1223 systems.}
\label{fig2}
\end{figure}

The crystals used for Raman measurements under hydrostatic pressure were prepared close to the optimal doping where the SC transition temperature $T_c$ is maximum. The Hg-1201 and Hg-1223 single crystals were synthesized and annealed following the method described in~\cite{Legros2019} and \cite{Loret2017} respectively. We sieved the batch immediately after annealing, in order to select crystals between \num{100} and \SI{200}{\micro\meter} in size. Samples with a naturally clean surface were directly selected. Their critical temperature $T_{c}$ have been determined from DC magnetization susceptibility measurements under classical zero field cooling (ZFC) on a set of crystals of the same batch. A PPMS magnetometer (Quantum Design) was used and a magnetic field of \num{10} Gauss was applied. The DC magnetization curves of the Hg-1201 and Hg-1223 single crystals at ambient pressure are displayed in Fig.~\ref{fig2} (a) and (b). The transition temperature \Tc and its width, $\Delta T_c$, were estimated by taking the maximum and the full width at half maximum of the peak of the first derivative of each DC magnetization curves, giving $T_{c}= 92 \pm 1~K $ and $T_{c}= 132\pm 2~K $ for Hg-1201 and Hg-1223 respectively. The two sets of samples are slightly under-doped and we name the ones selected for the Raman measurements under pressure (UD92K) for Hg-1201 and (UD132K) for Hg-1223 respectively. The $T_c$ values of the mercurate single crystals studied are reported in Fig.~\ref{fig2} (c) and (d).
%%%%%%%%%%%%%%%%%%%%%%%%%%%%%%%%%%%%%%%%%%%%%%%%%%%%%%%%%%%%%%%%%%%%%%%%%%%%%%%%%%%%%%%%%%%%%%%%%%%%%%%%%%%%%%%%
%%%%%%%%%%%%%%%%%%%%%%%%%%%%%%%%%%%%%%%%%%%%%%%%%%%%%%%%%%%%%%%%%%%%%%%%%%%%%%%%%%%%%%%%%%%%%%%%%%%%%%%%%%%%%%%%

\subsection{B. Polarized Raman experiments under hydrostatic pressure at low temperature}

\begin{figure}[ht!]
\includegraphics[scale=0.4]{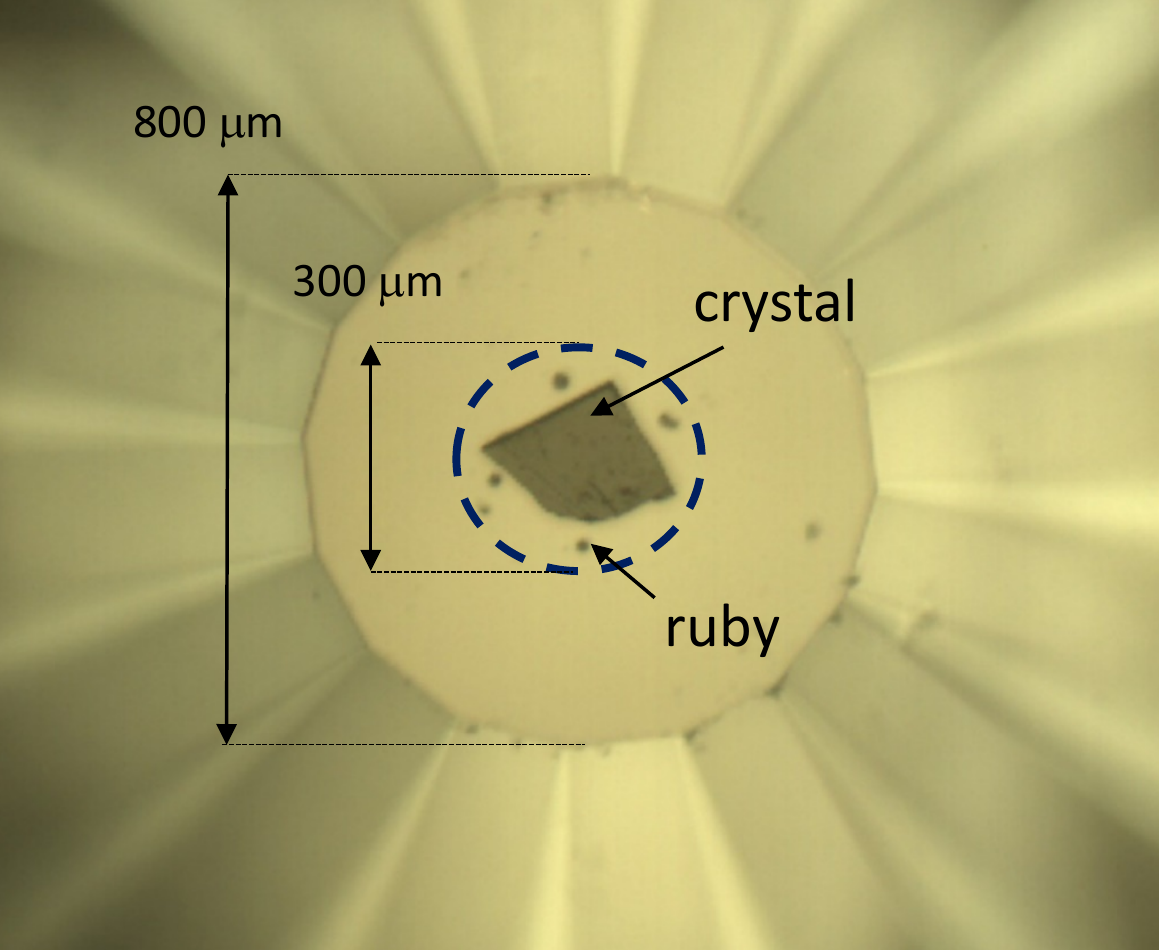}
\caption{Top view of the diamond anvil cell with diamonds of \SI{800}{\micro\meter} diameter before loading. The chamber between the diamonds is a cylindrical hole. After loading the chamber size is approximately \SI{300}{\micro\meter} (dashed circle).}
\label{fig3}
\end{figure}

\begin{figure}[ht!]
\includegraphics[scale=0.4]{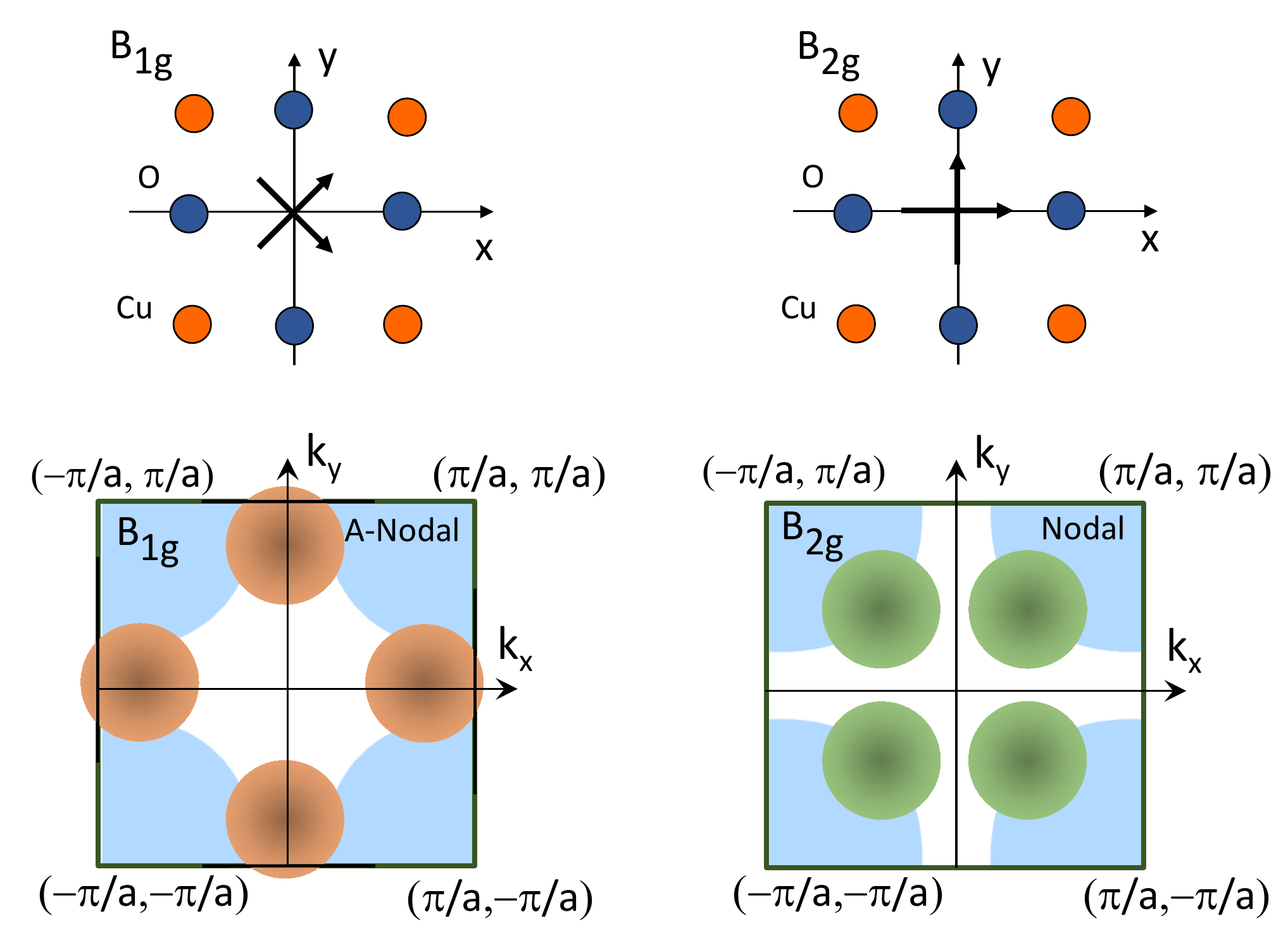}
\caption{Cross polarizations of the incoming and outgoing light (a) at 45 degrees from the Cu-O bond direction of the $\mathrm{CuO}_2$ plane corresponding to the \BAN geometry and (b) along the Cu-O bond direction corresponding to the \BN geometry. (c) The anti-nodal region probed by the \BAN geometry and (d) the nodal region probed by the \BN geometry in the first Brillouin zone.}
\label{fig4}
\end{figure}

We have performed two different runs of Raman measurements under pressure for studying the Hg-1201 and Hg-1223 compounds. Crystals were loaded inside a diamond anvil cell with diamonds of diameter \SI{800}{\micro\meter} designed to withstand up to \SI{10}{\giga\pascal}. The chamber between the diamonds is a cylindrical hole, cut by laser through a stainless steel gasket. Rubies were added in the cell to act as in-situ manometers through their fluorescence. Using several rubies allowed us to control the uniformity of the hydrostatic pressure in the chamber. After loading the chamber with ultra-pure helium and sealing it, the chamber size is approximately \SI{300}{\micro\meter} of diameter by \SI{50}{\micro\meter} of height (cf. Fig.~\ref{fig3}). 
%%%%%%%%%%%%%%%%%%%%%%%%%%%%%%%%%%%%%%%%%%%%%%%%%%%
%%%%%%%%%%%%%%%%%%%%%%%%%%%%%%%%%%%%%%%%%%%%%%%%%%%
The diamond anvil cell was then installed in a cryostat including a helium inlet allowing us to tune in-situ the pressure applied on the lower diamond, which indirectly changes the pressure in the chamber. Raman measurements were performed through the Boehler-designed upper diamond of the anvil, with a \SI{532}{\nano\meter} laser wavelength and \SI{4}{\milli\watt} of power (measured before going through the cryostat windows and the diamond). The spot size is around 40 microns in diameter with a slightly elliptical shape. According to our estimates, we can expect an upper limit for the laser heating to be around 1K/mW. Experiments were carried out using a JY-T64000 spectrometer in triple grating (1800 grooves/mm) configuration. The spectrometer is equipped with a nitrogen CCD detector. All the Raman spectra have been corrected for the Bose factor and the instrumental spectral response. They are thus proportional to the imaginary part of the Raman response function $\chi^{\prime \prime}(\omega,T)$. The Raman responses in the different symmetries are obtained from incoming and outgoing light polarizations. The B$_{1g}$ and B$_{2g}$ symmetries were obtained respectively from crossed polarizations of the incoming and outgoing light at 45 degrees and along the Cu-O bond direction of the $\mathrm{CuO}_2$ plane (cf. Fig.~\ref{fig4}). The (A$_{1g}$ + B$_{1g}$) geometry was got from parallel polarizations of the incoming and outgoing lights along the Cu-O bond direction. Raman scattering measurements at ambient pressure ($\approx~0$ GPa) were additionally performed with crystals of the same batches, outside of the diamond anvil cell, with the same experimental setup.

%%%%%%%%%%%%%%%%%%%%%%%%%%%%%%%%%%%%%%%%%%%%%%%%%%%
%%%%%%%%%%%%%%%%%%%%%%%%%%%%%%%%%%%%%%%%%%%%%%%%%%%

\section{III. Overview of the superconducting and normal Raman responses of Hg-1201 and Hg-1223 crystals at ambient pressure}

\begin{figure}[ht]
\includegraphics[scale=0.4]{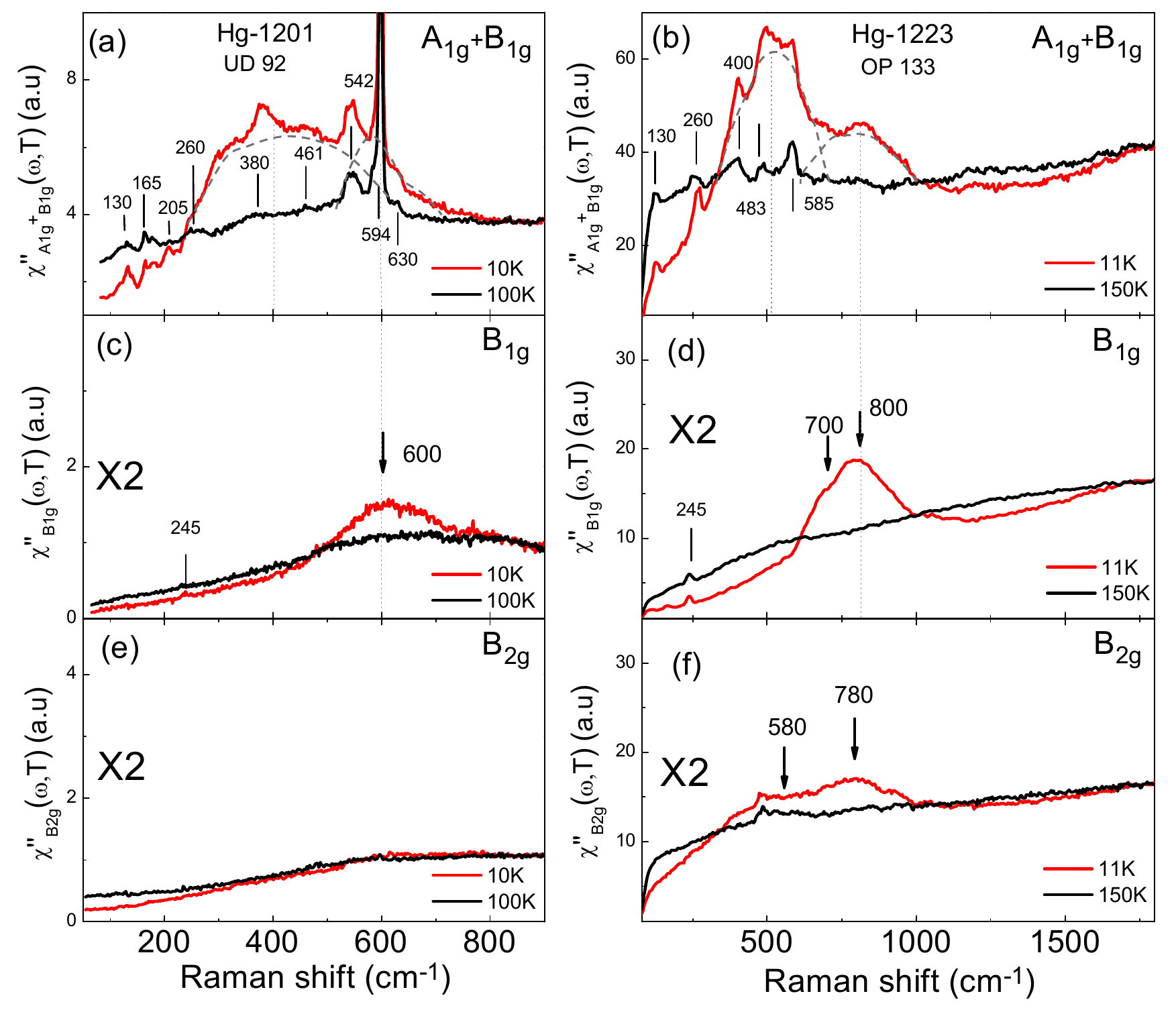}
\caption{Superconducting and normal Raman responses of the UD92K Hg-1201 and OP133K Hg-1223 in three distinct geometries at ambient pressure. The straight lines and numbers indicate the locations of the vibrational modes. Dashes lines indicate the A$_{1g}$ and \BAN contributions.The panel (f) contains a part of one spectrum in ref.\cite{Loret2019}.}
\label{fig:5}
\end{figure}

Our first objective is to disentangle the electronic excitations from the Raman-active optical phonons in order to study the electronic signatures of the superconducting phase of the Hg-1201 and Hg-1223 under hydrostatic pressure. The superconducting Raman responses at ambient pressure of a slightly under-doped (UD92K) Hg-1201 and an optimally doped (OP133K) Hg-1223 single crystals in three distinct geometries are reported in Fig.~\ref{fig:5}. They are made up of a broad electronic background superimposed by narrow peaks due to optical phonons. We focus first on the electronic background.

%%%%%%%%%%%%%%%%%%%%%%%%%%%%%%%%%%%%%%%%%%%%%%%%%%%%%%%%%%%%%%%%%%%%%%%%%%%%%%%%%%%%%%%%%%%%%%%%%%%%%%%%%%%%%%%%
%%%%%%%%%%%%%%%%%%%%%%%%%%%%%%%%%%%%%%%%%%%%%%%%%%%%%%%%%%%%%%%%%%%%%%%%%%%%%%%%%%%%%%%%%%%%%%%%%%%%%%%%%%%%%%%%
 
\subsection{A. Electronic part of the Raman response}

Electronic Raman scattering is a particularly useful probe for studying the cuprates because we can select distinct parts of the BZ, the anti-nodal and nodal regions well known to have quite different electronic properties \cite{Norman03}. In the \BAN geometry the Raman form factor is $( \cos k_x - \cos k_y)^2$ and it predominantly probes the anti-nodal region where the superconducting gap and the pseudogap are maximal. Here ${\bf k}$ is the wave vector of the excited electron. Likewise, in the \BN geometry the Raman form factor is $\sin^2 k_x \sin^2 k_y$ and it probes mostly the nodal region where the superconducting gap and the pseudogap are minimal. In the A$_{1g}$ geometry, the Raman form factor is more isotropic, with no symmetry-imposed nodes. Experimentally, we cannot access the pure $A_{1g}$ component using linear polarizations, as it is always associated with either a B$_{1g}$ or B$_{2g}$ component. The (A$_{1g}$ + B$_{1g}$) SC Raman spectra of (UD92K) Hg-1201 and (OP133K) Hg-1223 single crystals (red curve at 10 K in Fig.~\ref{fig:5} (a) and (b)) show an extended hump in energy made up of two broad peaks, the A$_{1g}$ and \BAN peaks. The A$_{1g}$ and \BAN peaks are respectively located around 400 \cm and 600 \cm in the Hg-1201 spectrum and centered around 500 \cm and 800 \cm in the  Hg-1223 spectrum. These features are indicated by gray dashed lines in the spectra and extensively studied in previous works \cite{Gallais04,Guyard2008a,LeTacon2006,Benhabib2015}. The \BAN SC peak alone are displayed in Fig.~\ref{fig:5} (c) and (d). It corresponds to the pairs breaking peak related to the maximum amplitude of the $d-$ wave SC gap opening. No clear SC peak is detected in the \BN Raman spectra of Hg-1201 and a relatively weak peak (close to 780 \cm) compared to the \BAN peak is detected in the \BN Raman spectra of Hg-1223 (cf. Fig.~\ref{fig:5} (e) and (f)). This is expected since the SC gap vanishes out in the nodal regions probed predominantly by the \BN geometry. In particular, the remarkably flat \BN SC Raman response of the (UD92K) Hg-1201 will be exploited later for the electronic Raman measurements under pressure. Notice that \BN spectra of Hg-1223 contains an extra electronic contribution (around 580 \cm) stemming from the charge density wave order as already discussed in our previous investigations \cite{Loret2019,Loret2020}.

%%%%%%%%%%%%%%%%%%%%%%%%%%%%%%%%%%%%%%%%%%%%%%%%%%%%%%%%%%%%%%%%%%%%%%%%%%%%%%%%%%%%%%%%%%%%%%%%%%%%%%%%%%%%%%%%
\begin{figure}[ht]
\includegraphics[scale=0.4]{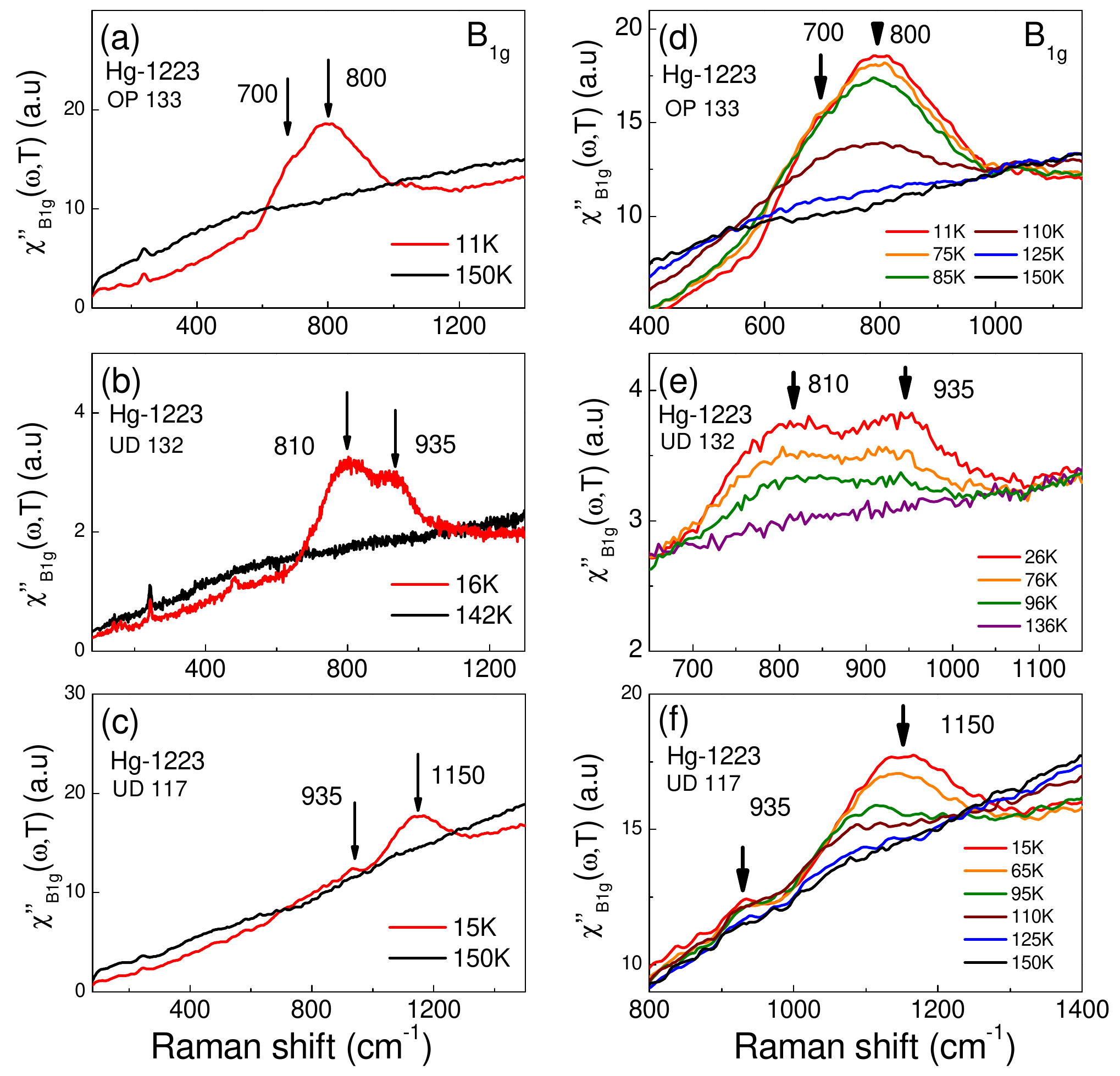}
\caption{Double structure of the \BAN superconducting gap of the Hg-1223 (a)-(c) as a function of doping (d)-(f) as function of temperature for each doping. The panel (f) contains a part of one spectrum in ref.\cite{Loret2020}.}
\label{fig:6}
\end{figure}
%%%%%%%%%%%%%%%%%%%%%%%%%%%%%%%%%%%%%%%%%%%%%%%%%%%%%%%%%%%%%%%%%%%%%%%%%%%%%%%%%%%%%%%%%%%%%%%%%%%%%%%%%%%%%%%%
\par
A more detailed analysis reveals that the \BAN superconducting peak in the Raman spectrum of the three-layer Hg-1223 cuprate presents a shoulder (around 700 \cm) on its left side that we do not observe in the Raman spectrum of the one-layer Hg-1201 cuprate (see  panels (c) and (d) Fig.~\ref{fig:5}). To have a better understanding of its origin, we followed the evolution of this shoulder as a function of doping. The shoulder that we observe close to the optimal doping turns into a double peak with under-doping. This is displayed in Fig.~\ref{fig:6} (a)-(c). 
The frequency difference between the two peaks increases with under-doping. These two peaks could be related to two superconducting gaps induced either by an inter, or by an intra-unit cell doping inhomogeneity of Hg-1223. Our experimental findings tend towards an intra-unit cell doping inhomogeneity due to a charge carrier concentration imbalance between the inner and the outer $\mathrm{CuO}_2$ planes of one single unit cell. Indeed, the Raman spectra measured on the Hg-1223 crystal for a given doping level, are the same whatever the location of the laser spot on the crystal surface. This means there is no trace, at least at the scale of few ten microns, of inter cell inhomogeneity of the oxygen doping. Secondly, we always detect only two distinct superconducting peaks (cf. Fig.~\ref{fig:6} (a)-(c)) and not more, which seems inconsistent with a distribution of spatial oxygen content over the area illuminated by the laser spot. The observation of only two peaks which deviate from each other in energy as the doping decreases is rather in favor of the existence of two SC gaps linked to the inner and outer $\mathrm{CuO}_2$ planes which would have different carriers concentrations. This is in agreement with previous works on Bi-2223 three layers cuprate \cite{Ideta2010,Vincini2018}. Thirdly, the two SC peaks detected in the Raman spectra seem to disappear simultaneously (within our temperature accuracy) as the temperature is raised (cf. Fig.~\ref{fig:6}(d)-(f)). This suggests that the two gaps are interconnected as it should be the case if they are originate from the inner and outer plane of the same unit cell and obviously not if the two gaps come from distinct regions of the crystal with different oxygen contents. The inner plane being more distant from the reservoir blocks by which the charge transfer takes place, earlier studies have concluded that the charge carriers concentration of the $\mathrm{CuO}_2$ outer plane is higher than the one of the inner plane in the under-doped Hg-1223 \cite{Kotegawa2002,Julien1996,Mukuda2012,Mukuda2016,Mito2017,Iwai2014}. The \BAN superconducting gap energy is known to be larger upon reduced doping in Hg-1223 \cite{Loret2019,Loret2020}. We can thus assign the high energy peak to the SC gap related to the inner plane and the lower one to the SC gap of the outer plane.

\subsection{B. Phononic part of the Raman response}

In Fig.~\ref{fig:5} (a) and (b)) few narrow peaks (marked by straight lines) are superimposed to the (A$_{1g}$ + B$_{1g}$) electronic background.  They correspond to Raman active phonons associated with the Hg-1201 and Hg-1223 structures as well as parasitic phases. Overall, we detect three types of phonons. The first type is associated with the vibrational modes of the ideal stoichiometric Hg-1201 and Hg-1223 structure that belong to the $D^1_{4h}$space group. We then expect $2A_{1g}+ 2E_{2g}$ and $5A_{1g}+ 1B_{1g}+ 6 E_{g}$ pristine Raman active even (gerade) modes for the Hg-1201 and Hg-1223 structure respectively. The second kind of vibrational modes is associated with defect stemming from symmetry breaking induced by insertion of oxygen atoms in the $\mathrm{HgO}$ layers which makes Raman active some odd (ungerade) modes. At last, the third kind of vibrational modes come from parasitic phases which are deposited on the crystal's surface. They stem from residual oxides of synthesis precursor phases that subsists even in the cleanest crystal surfaces we could select. A summary of these three kinds of phonons observed in Hg-1201 and Hg-1223 structures is reported in Table 1. \\

%%%%%%%%%%%%%%%%%%%%%%%%%%%%%%%%%%%%%%%%%%%%%%%%%%%%%%%%%%%%%%%%%%%%%%%%%%%%%%%%%%%%%%%%%%%%%%%%%%%%%%%%%%%%%%
\begin{table*}
\begin{center}
\begin{tabular}{|*{7}{c|}}
		\hline
		\hline
 \textbf{ Hg-1201}  &   &  &   &   &   &  \\
    \hline
  Pristine mode  &  &165\newline (Ba)&   &   &   & 594 (O2)\\
		\hline
	Defect mode   &130 (Cu-Ba-Hg')&  &260 (O1')&461 (O1)&542 (O3)&\\
		%\hline
	%Parasitic mode &  &207 (Hg-Ba-O)&310 (Ba-Cu-O)&370 (Hg-O)&  &620 (Ba-Cu-O)\\
    \hline
		\hline
%\end{tabular}
%\end{center}
%%%%%%%%%%%%%%%%%%%%%%%%%%%%%%%%%%%%%%%%%%%%%%%%%%%%%%%%%%%%%%%%%%%%%%%%%%%%%%%%%%%%%%%%%%%%%%%%%%%%%%%%%%%%%%%
%%%%%%%%%%%%%%%%%%%%%%%%%%%%%%%%%%%%%%%%%%%%%%%%%%%%%%%%%%%%%%%%%%%%%%%%%%%%%%%%%%%%%%%%%%%%%%%%%%%%%%%%%%%%%%%
%\begin{center}
%\begin{tabular}{|*{7}{c|}}
  \textbf{Hg-1223}  &   &  &   &   &   &  \\
    \hline
  Pristine mode  & 120 (Ba) &260 (O4')& 483 (O4)&   &   & 585 (O2)\\
		\hline
	Defect mode   &130 (Cu-Ba-Hg')&  & &400 (O1)& &\\
		\hline
		\hline
	Parasitic mode &  &205      &310      &380    & 570 		&625 \\
	               &  &(Hg-Ba-O)&(Ca-Cu-O)& (Hg-O)&(Ba-Cu-O)&(Ba-Cu-O)\\
		\hline
		\hline

\end{tabular}
\end{center}
\caption{Enumeration of the A$_{1g}$ + B$_{1g}$ vibrational modes (\cm unit) detected in the Raman spectra of Hg-1201 and Hg-1223 single crystals. The pristine modes are all associated with vertical motions along the $c$-axis. The prime denotes counter phase displacement. The lattice dynamical calculation can be found in ref.\cite{Popov1995} and previous attempts for assigning the Raman active modes in Hg-1201 and Hg-1223 structures can be found in ref.\cite{Krantz1994,Sacuto1996,Zhou1996,Wang2018}. Some of defect modes and parasitic phases can be found in earlier works \cite{Krantz1994,Sacuto1996,Zhou1996,Wang2018,Ren1993,Hur1993,Sacuto2000b}. The (Ba-Cu-O), (Ca-Cu-O) and (Hg-O) compositions are generic terms which can involve different parasitic phases namely: BaCuO$_2$, Ba$_2$CuO$_3$, Ba$_2$Cu$_3$O$_{5/6}$, Ca$_2$CuO$_3$ and HgBaO$_2$. The row of parasitic modes includes those of the Hg-1201 and Hg-1223 structures.}
\end{table*}
%%%%%%%%%%%%%%%%%%%%%%%%%%%%%%%%%%%%%%%%%%%%%%%%%%%%%%%%%%%%%%%%%%%%%%%%%%%%%%%%%%%%%%%%%%%%%%%%%%%%%%%%%%%%%%%

\section{IV. Overview of the superconducting and normal Raman responses of Hg-1201 and Hg-1223 crystals under hydrostatic pressure}

\begin{figure}[ht]
\includegraphics[scale=0.4]{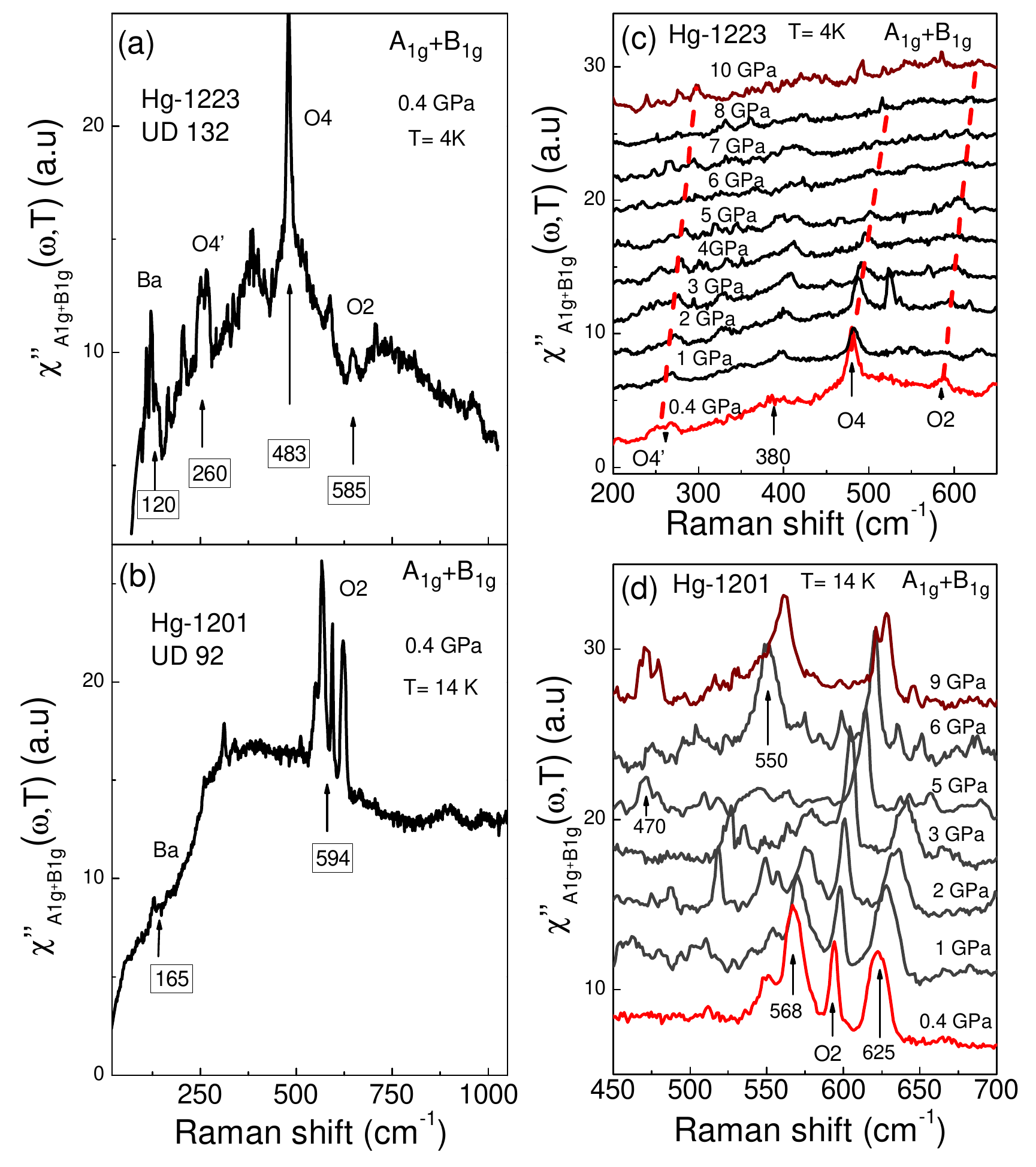}
\caption{Superconducting Raman responses in the $A_{1g}+B_{1g}$ geometry of (a) the Hg-1201 and (b) Hg-1223 crystal at 0.4 GPa. Zoom on the frequency range of the $A_{1g}+B_{1g}$ Raman spectra of (c)  Hg-1201 and (d) Hg-1223 compounds under hydrostatic pressures. The framed frequencies mark the pristine oxygen modes (see Table 1). The dotted lines are guides for the eyes.}
\label{fig:7}
\end{figure}

The $A_{1g}+B_{1g}$ Raman superconducting responses of (UD92K) Hg-1201 and (UD132K) Hg-1223 single crystals measured inside the anvil cell at 0.4 GPa are reported in Fig.~\ref{fig:7}. We have nearly the same spectra as those of Fig.~\ref{fig:5} but with a poorer signal-to-noise ratio and few additionnal parasitic phonons (at 570 and 620 \cm). We see an asymmetric hump with a maximum around 400 \cm in the Hg-1201 which is made of two components (cf. Fig.~\ref{fig:5}). The Hg-1223 spectrum exhibits two broad peaks, one centered around 480 \cm and the second one around 760 \cm which are assigned the $A_{1g}$ and $B_{1g}$ components of the spectrum (cf. Fig.~\ref{fig:5}). On the top of these broad electronic peaks few narrow peaks associated with vibrational modes are detected. The pristine $A_{1g}+B_{1g}$ phonons related to the oxygen motions of the Hg-1223 and Hg-1201 structures are indicated by black black arrows in panel (a) and (b) of Fig.~\ref{fig:7}. %Sometimes some extra modes appear at high pressure in the other geometries : around 400 \cm-1 in Hg-1223 or close to 260, 470 et 550 \cm in Hg-1201. They are stemming from defect modes (cf. Table 1) and induced by mechanical stress as explained below. 

%%%%%%%%%%%%%%%%%%%%%%%%%%%%%%%%%%%%%%%%%%%%%%%%%%%%%%%%%%%%%%%%%%%%%%%%%%%%%%%%%%%%%%%%%%%%%%%%%%%%%%%%%%%%%%%%
%%%%%%%%%%%%%%%%%%%%%%%%%%%%%%%%%%%%%%%%%%%%%%%%%%%%%%%%%%%%%%%%%%%%%%%%%%%%%%%%%%%%%%%%%%%%%%%%%%%%%%%%%%%%%%%%

\subsection{A. Study on the oxygen pristine modes of Hg-1223 and Hg-1201 with pressure.}

\begin{figure}[ht]
\includegraphics[scale=0.35]{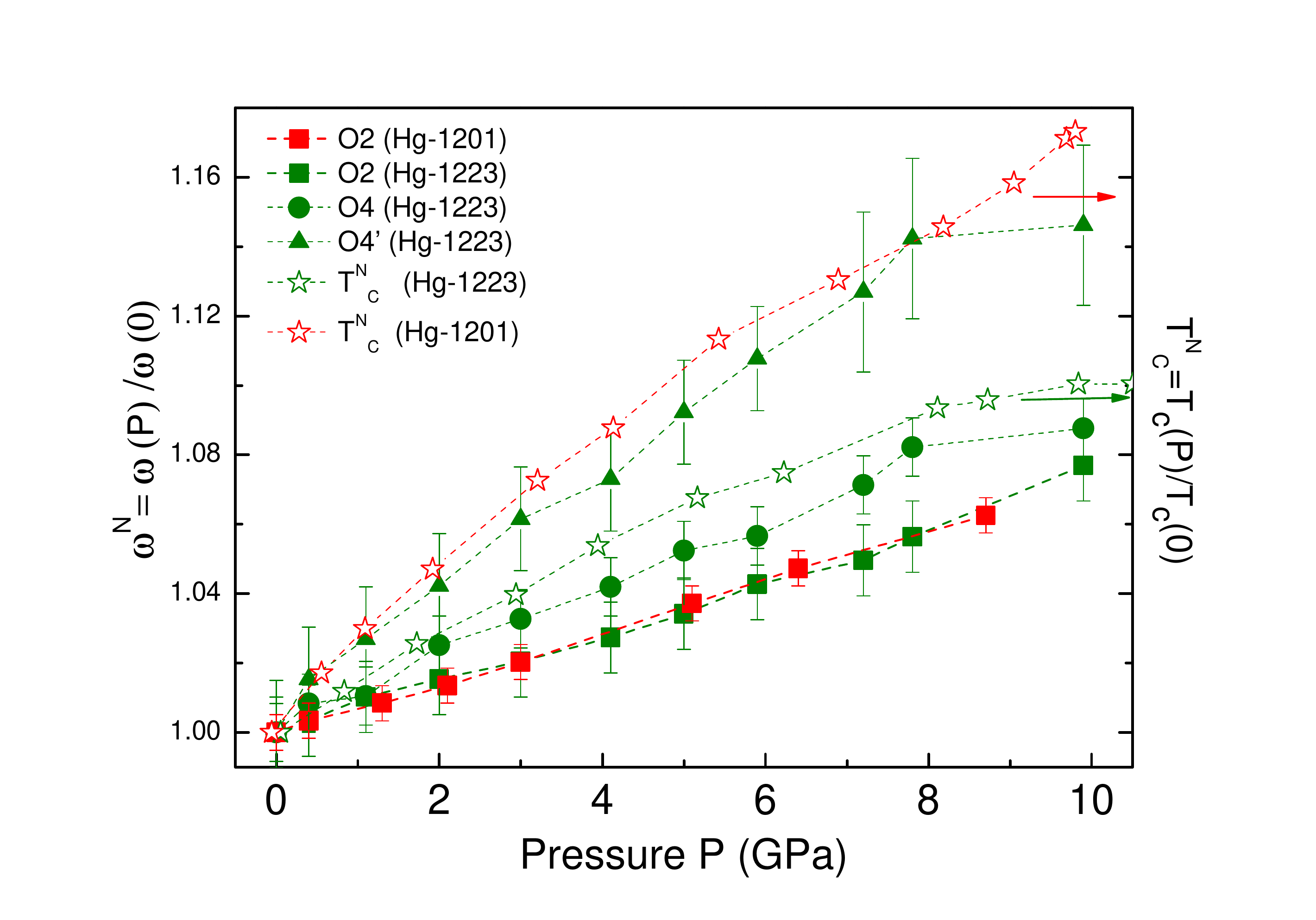}
\caption{The normalized frequencies of the oxygen vibrational modes of Hg-1223 and Hg-1201 as a function of pressure $P$. The normalized frequency is obtained by dividing the frequency at pressure $P$ by this value at ambient pressure. The normalized critical temperature $\Tc^N=T_c(P)/T_c(0)$ is also reported. The $T_c(P)$ values come from ref. \cite{Antipov2002,Yamamoto2015}.}
\label{fig:8}
\end{figure}

We displayed in Fig.~\ref{fig:7} (c) and (d) a spectral range zoom of the Raman spectra of panels (a) and (b). The frequencies of the oxygen pristine modes of the Hg-1223 at 260, 483 and 585 \cm (see Table 1) increase with pressure (see red dotted lines in panel (c)). The  feature centered around 380 \cm  is probably a parasitic mode (see Table 1). In the Hg-1201 spectra (panel (d)), the apical oxygen mode at 594 \cm (framed) increases with pressure. It is located in between two parasitic modes at 570 and 625 \cm (see Table 1) whose the full width at half maximum is three time larger. These two peaks disappear at high pressure presumably because the structure of the parasitic phases evolves under pressure. We can also notice that two peaks at 470 and 550 \cm appear with pressure (see arrows). They are likely defect modes that come out due to mechanical stress which induce a redistribution of mobile oxygen in the $\mathrm{HgO}$ layer \cite{Lorenz2005}. They correspond to the defect modes located at 461 and 542 cm-1 (reported in Table 1) and whose frequencies increase with pressure. The normalized frequencies of the oxygen pristine modes of the Hg-1223 and Hg-1201 structure as a function of pressure, $P$, are displayed in Fig.~\ref{fig:8} 

%%%%%%%%%%%%%%%%%%%%%%%%%%%%%%%%%%%%%%%%%%%%%%%%%%%%%%%%%%%%%%%%%%%%%%%%%%%%%%%%%%%%%%%%%%%%%%%%%%%%%%%%%%%%%%%%
%%%%%%%%%%%%%%%%%%%%%%%%%%%%%%%%%%%%%%%%%%%%%%%%%%%%%%%%%%%%%%%%%%%%%%%%%%%%%%%%%%%%%%%%%%%%%%%%%%%%%%%%%%%%%%%%

A clear trend emerges: the normalized frequencies of the oxygen vibrational modes of the Hg-1223 and Hg-1201 compounds increase almost linearly with pressure up to \SI{7}{\giga\pascal}. This behavior is usually expected with pressure \cite{Goncharov1992,Goncharov2003,Cuk2008,Aupiais2018}. The $\mathrm{CuO}_2$ layer of Hg-1201 being a symmetry plane (cf. Fig.~\ref{fig1}), the vibrational modes related to the (O4) and (O4') oxygen motion in the $\mathrm{CuO}_2$ are non Raman active in this structure. Therefore, we will focus on the (O2) vibrational mode which is present both in the Hg-1201 and Hg-1223 structure. Remarkably, the slopes ($\frac{d\omega^N}{dP}$) of the apical oxygen (O2) phonon over the entire pressure range are nearly the same in the Hg-1201 and Hg-1223 structures (see red and green full squares). On the contrary, the slopes of the normalized \Tc, $\frac{d\Tc^N}{dP}$ associated with the Hg-1201 and Hg-1223 compounds are different (see red and green open stars). This means that the pressure evolution of \Tc in these structures cannot be attributed to the apical oxygen mode dynamics alone. The number of $\mathrm{CuO}_2$ planes and their respective charge carrier concentration must also be taken into account. In particular, previous investigations \cite{Trokiner1991,Julien1996,Kotegawa2002,Mukuda2012,Iwai2014,Mito2017} revealed that the carriers concentration imbalance between the $\mathrm{CuO}_2$ of multilayer superconductors such as Hg-1223 was limiting the increase of \Tc. This can be visualized in Fig.~\ref{fig:8}. The $\Tc^N$ slope of Hg-1201 is much higher than the Hg-1223 likely due to the difficulty to perform an efficient charge transfer by pressure to the inner plane of Hg-1223 as it will be shown in the next section. \\
At this stage, it is interesting to make a comparison between the effects of pressure (P) and doping (p) on the dynamic of the apical oxygen mode in the Hg-1201 and Hg-1223 structure. The apical oxygen atom (O2) is the bridge between the $\mathrm{HgO}$ and the $\mathrm{CuO}_2$ planes (see Fig.~\ref{fig1}) and its motion is along the $c$ axis. The apical oxygen mode hardens with pressure (cf. Fig.~\ref{fig:8}) while it softens with doping \cite{Sacuto1996,Zhou1996,Legros2019,Loret2017}. Its softening is due to its coupling with charge carriers which increases with doping \cite{Balkanski1975,Slakey1989,Friedl1990} and is a common feature of many cuprates such as Hg-1201, Y-123 and Hg-1223 \cite{Legros2019,Thomsen1988,Sacuto1993,Bakr2013,Sacuto1996,Loret2017}. On the other hand, its hardening with pressure is likely due to the inter-plane contraction along the $c$ axis \cite{Eggert1994,Mito2017} which is much stronger with pressure than with doping. A way to compare the $c$ axis contraction under pressure and doping is to evaluate it as a function of the \Tc change. According to the refs.\cite{Fukuoka1997,Antipov2002,Eggert1994,Mito2017,Loret2017,Legros2019}, the $\frac{\Delta c}{\Delta \Tc (P)}$ ratio is approximately one order of magnitude larger than the $\frac{\Delta c}{\Delta \Tc (p)}$. Our estimate is $\frac{\Delta c}{\Delta \Tc(P)}\approx 10^{-2}$ \AA K$^{-1}$ and $\frac{\Delta c}{\Delta \Tc (p)}< 10^{-3}$ \AA K$^{-1}$. This justifies why the apical oxygen mode hardens under pressure even if applying pressure favor the charge transfer and an increase of the charge carriers which softens the apical oxygen mode in the case of chemical doping \cite{Ohta1991,Yamamoto2015,Mito2017}. Despite these contrasting trends in the dynamics of the apical mode with pressure and doping, understanding the roles that pressure and doping play on \Tc remains a major issue. We can cite two emblematic experimental facts that show their effects are different on \Tc. First, applying a pressure on an under-doped Hg-1212 or Hg-1223 compound allows us to reach a much higher \Tc than that which can be obtained by doping \cite{Yamamoto2015}. Secondly, starting from optimally doped Hg-1201 and Hg-1223, \Tc decreases with doping (over-doped regime) while it increases with pressure \cite{Antipov2002,Chu1993}. Consequently, the effects of pressure and doping on \Tc are complex and additional parameters to the dynamics of the apical mode have to be considered such as the shortening of the oxygen bond lengths in the $\mathrm{CuO}_2$ plane according to ref. \cite{Pavarini2001}.

\subsection{B. Study on the bare electronic superconducting Raman response of Hg-1223 and Hg-1201 with pressure.}

We focus now on the superconducting electronic Raman response under pressure. This is achieved by subtracting the most intense phonons after fitting them by lorentzian profiles \cite{LeTacon2007}. The bare electronic Raman responses in \BAN and \BN geometries of Hg-1223 and Hg-1201 free of phonons are reported in Fig.~\ref{fig:9}. In Appendix A are shown the raw data with phonons.

%%%%%%%%%%%%%%%%%%%%%%%%%%%%%%%%%%%%%%%%%%%%%%%%%%%%%%%%%%%%%%%%%%%%%%%%%%%%%%%%%%%%%%%%%%%%%%%%%%%%%%%%%%%%%%%%
\begin{figure}[ht]
\includegraphics[scale=0.4]{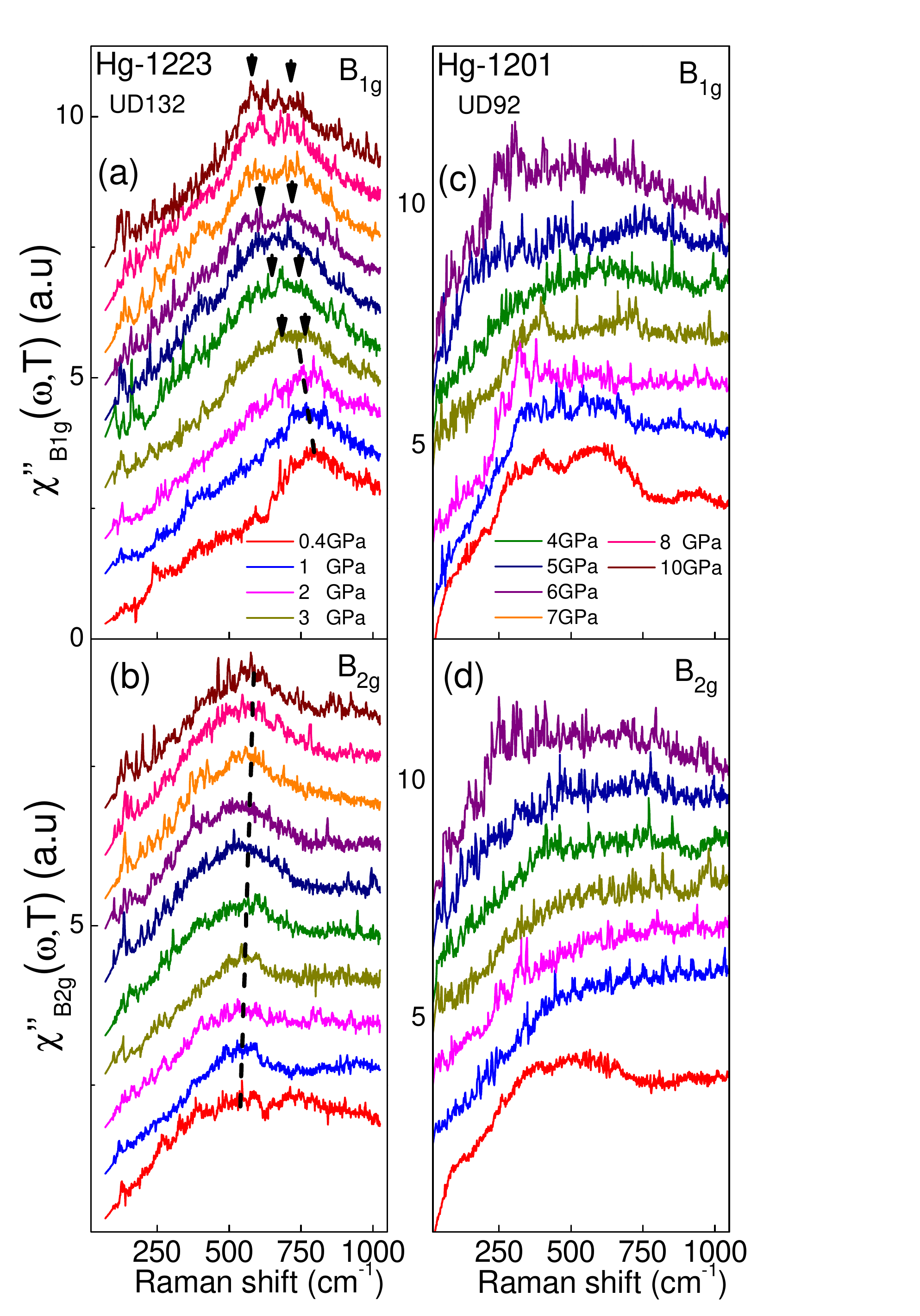}
\caption{$B_{1g}$ and $B_{2g}$ Raman responses free of phonons under hydrostatic pressures of Hg-1223 (a)-(b) and Hg-1201 (c)-(d). The sets of Hg-1223 and Hg-1201 data were obtained at $\SI{4}{\kelvin}$ and $\SI{14}{\kelvin}$ respectively.}
\label{fig:9}
\end{figure}
%%%%%%%%%%%%%%%%%%%%%%%%%%%%%%%%%%%%%%%%%%%%%%%%%%%%%%%%%%%%%%%%%%%%%%%%%%%%%%%%%%%%%%%%%%%%%%%%%%%%%%%%%%%%%%%%
 
\begin{figure}[ht]
\includegraphics[scale=0.35]{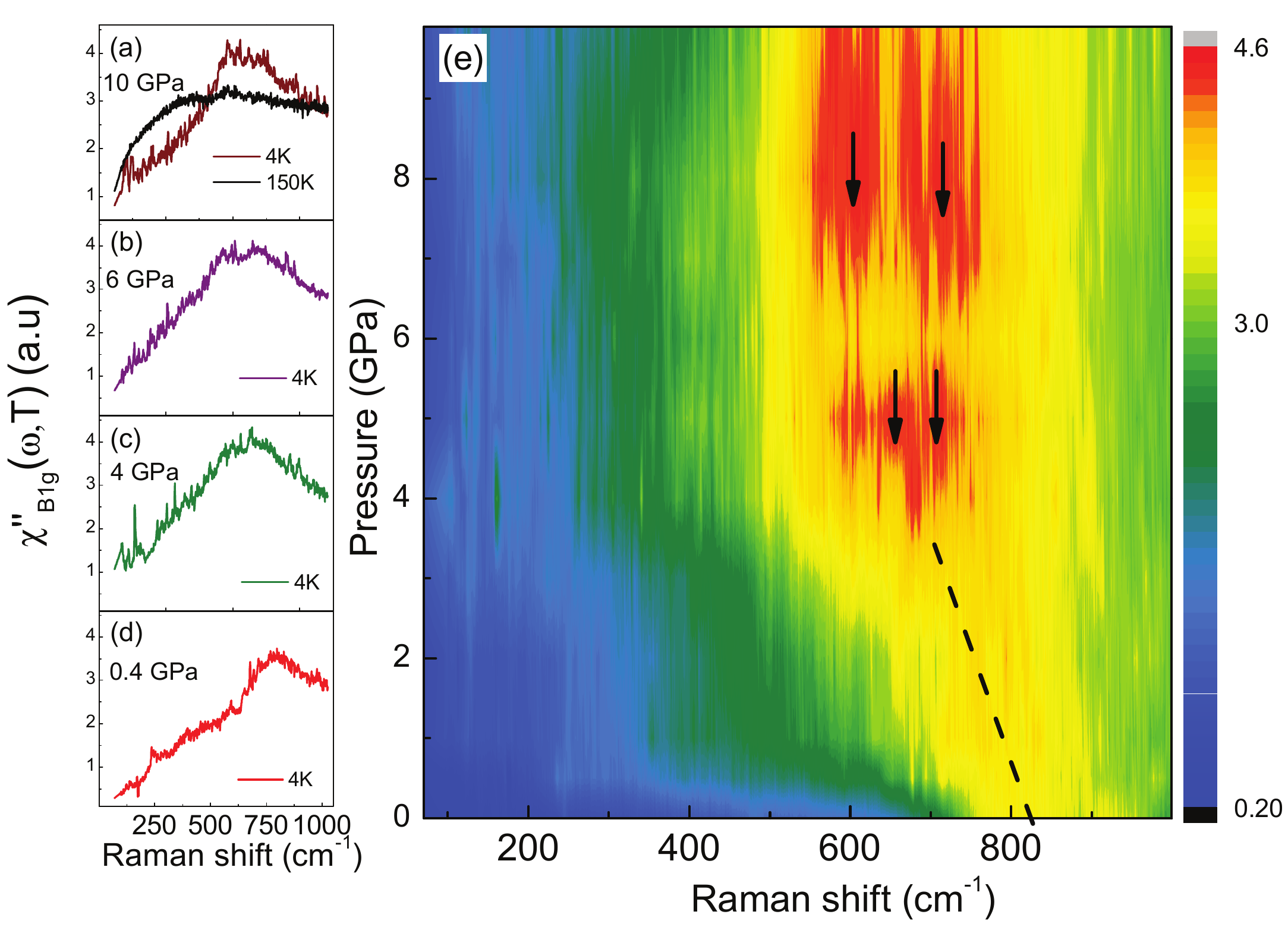}
\caption{(Color online) $B_{1g}$ Raman response in Hg-1223 under pressure at $T = \SI{4}{\kelvin}$. (a)-(d) Electronic Raman response featuring the superconducting peak at various pressures. At maximum \SI{10}{\giga\pascal} pressures, the normal state response above $T_c$ is also displayed in order to underlined the superconducting peak. (e) Contour plot highlighting the evolution of the $B_{1g}$ spectral weight as a function of pressure. The dashed lines are guides for the eyes.}
\label{fig:10}
\end{figure}

\begin{figure}[ht]
\includegraphics[scale=0.35]{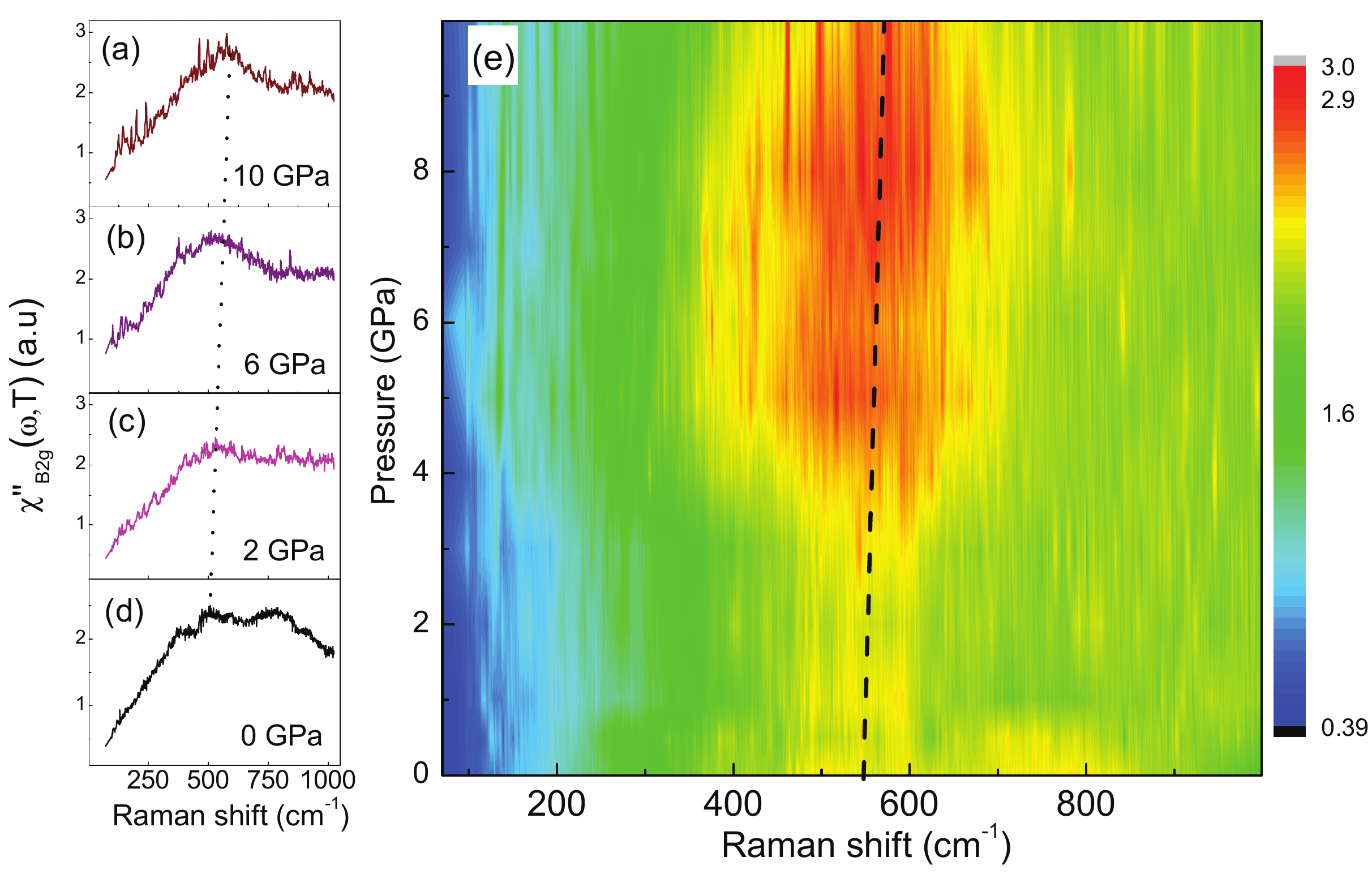}
\caption{(Color online) $B_{2g}$ Raman response in Hg-1223 under pressure at $T = \SI{4}{\kelvin}$. (a)-(d) Electronic Raman response featuring the superconducting peak. (e) Contour plot highlighting the evolution of the $B_{2g}$ spectral weight as a function of pressure. The dashed line is a guide for the eyes.}
\label{fig:11}
\end{figure}

\begin{figure}[ht]
\includegraphics[scale=0.4]{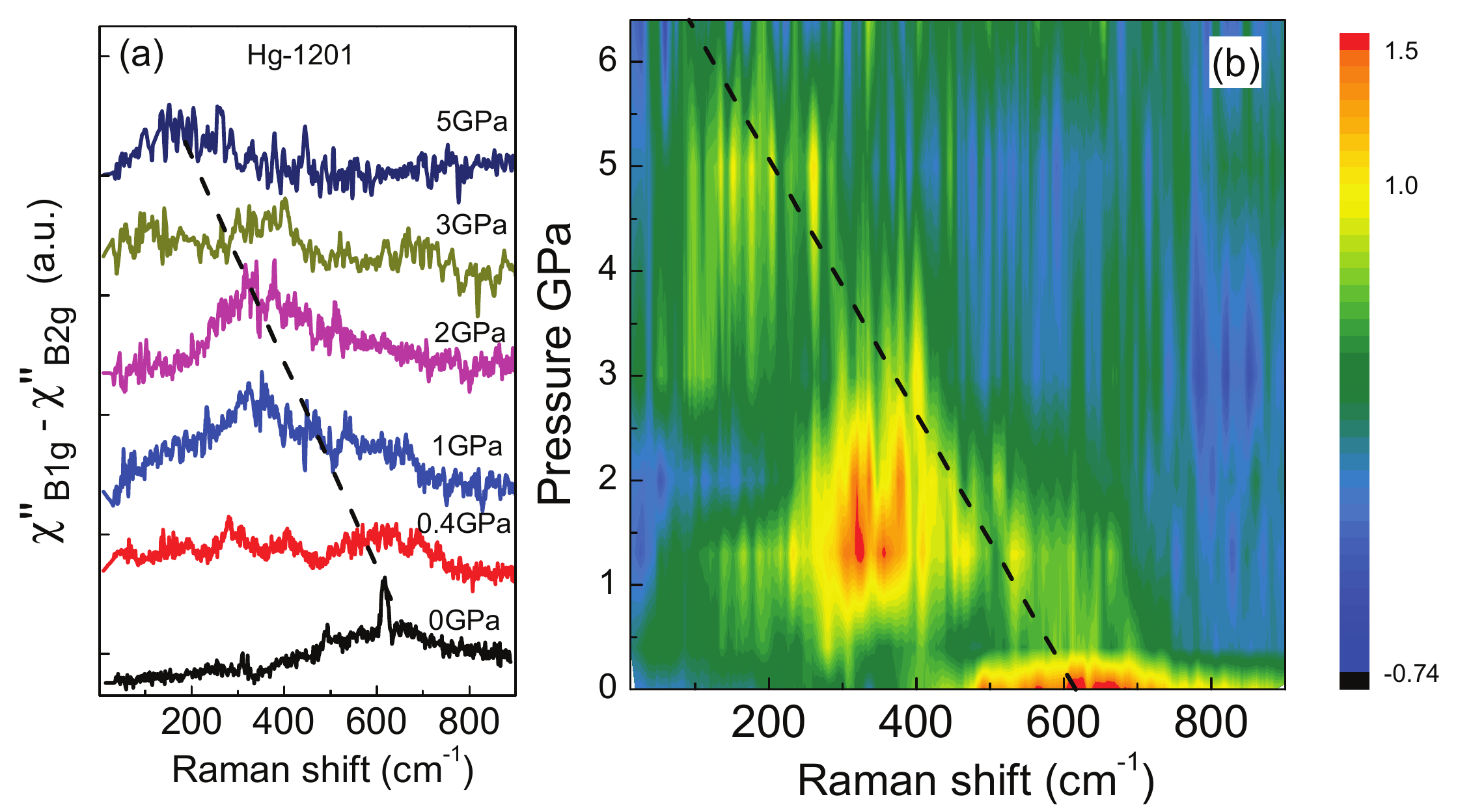}
\caption{(Color online) Following the \BAN SC gap in Hg-1201. (a) Difference between the \BAN and the \BN Raman response (the latter one serving as a reference background), as a function of pressure. (b) Contour plot highlighting the evolution of the $B_{1g}$ spectral weight with pressure. The dashed line is a guide for the eyes.}
\label{fig:12}
\end{figure}

\begin{figure*}[ht]
\includegraphics[scale=0.5]{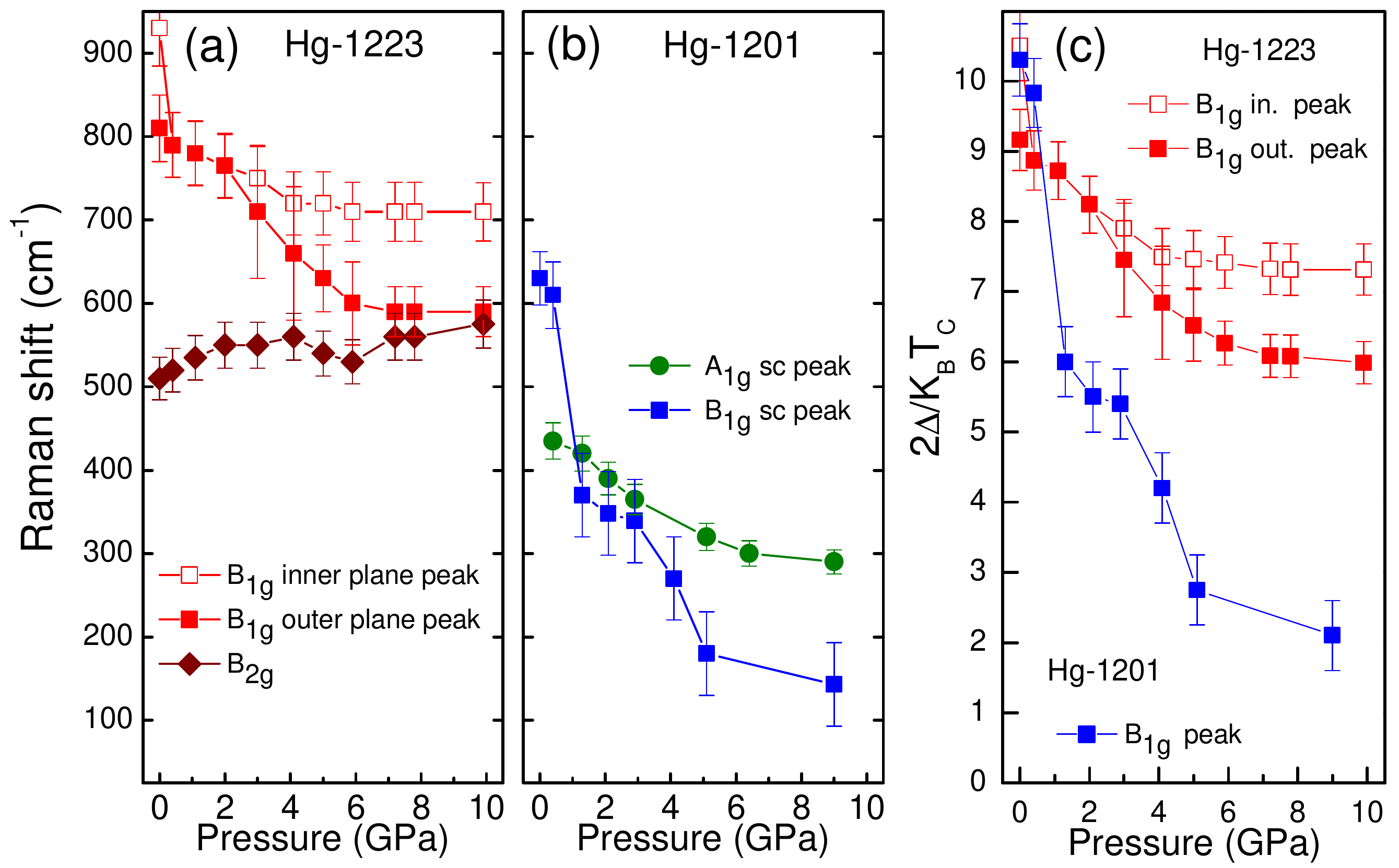}
\caption{(Color online) Frequency evolution of the characteristic superconducting peaks in the Raman spectra of Hg-1223 and Hg-1201 compounds.}
\label{fig:13}
\end{figure*}

The B$_{1g}$ electronic response of the UD132 Hg-1223 compound (cf.Fig.~\ref{fig:9} (a)) exhibits a hump which decreases in energy with pressure. This hump is the SC pair breaking peak and only appears below \Tc as shown in the Raman spectra at low and high pressure in  Fig.~\ref{fig:6} (b) and Fig.~\ref{fig:10} (a) respectively. The pair breaking peak centered around $2\Delta \approx$ 800 \cm at \SI{0.4} {\giga\pascal} has probably a double component like in Fig.~\ref{fig:6} (b), but the lower signal-to-noise ratio and an additional background to the spectra measured inside the cell likely prevents us from detecting the weaker high energy shoulder (see Appendix B). However, as the pressure increases, the width of the pair breaking peak increases and the splitting appears (see black arrows in Fig.~\ref{fig:9} (a)). This double component has already been identified (in section III) as the two gaps associated with the inner and outer CuO$_2$ planes of the Hg-1223 structure when these two planes do not have the same charge carrier concentration. As the pressure increases up, the charge carrier concentration imbalance is accentuated between the inner and outer planes. This makes it possible to detect again (but this time inside the cell) the splitting of the hump related to two superconducting gaps. This can be seen by the two red trails in the color map which correspond to the two SC peaks, see Fig.~\ref{fig:10} (e).
%%%%%%%%%%%%%%%%%%%%%%%%%%%%%%%%%%%%%%%%%%%%%%%%%%%%%%%%%%%%%%%%%%%%%%%%%%%%%%%%%%%%%%%%%%%%%%%%%%%%%%%%%%%%%%%%
%%%%%%%%%%%%%%%%%%%%%%%%%%%%%%%%%%%%%%%%%%%%%%%%%%%%%%%%%%%%%%%%%%%%%%%%%%%%%%%%%%%%%%%%%%%%%%%%%%%%%%%%%%%%%%%%
The \BN Raman spectra of Hg-1223 under pressure and low temperature presents also a hump associated with a SC peak (see Fig.~\ref{fig:9} (b)). It corresponds to the weaker SC gap feature associated with the nodal region (see section III). This hump is more and more visible as pressure increases and its frequency slightly increases as the pressure increases. This is highlighted in the colour map  of Fig.~\ref{fig:11} (e).\\
%%%%%%%%%%%%%%%%%%%%%%%%%%%%%%%%%%%%%%%%%%%%%%%%%%%%%%%%%%%%%%%%%%%%%%%%%%%%%%%%%%%%%%%%%%%%%%%%%%%%%%%%%%%%%%%%
%%%%%%%%%%%%%%%%%%%%%%%%%%%%%%%%%%%%%%%%%%%%%%%%%%%%%%%%%%%%%%%%%%%%%%%%%%%%%%%%%%%%%%%%%%%%%%%%%%%%%%%%%%%%%%%%
We focus now on the \BAN SC Raman response function of the Hg-1201 compound in Fig.~\ref{fig:9} (c). The \BAN Raman response at 0.4 GPa (red curve) exhibits an unexpected broad peak centered around 380 \cm in addition to the expected \BAN pair breaking peak at 600 \cm (cf.Fig.~\ref{fig:5} (c)). This extra peak is due to a leakage of an $A_{1g}$ contribution to the \BAN Raman response. It likely comes from an increase in the birefringence of the pressurized diamonds. The \BAN and \BN Raman spectra having been measured one after the other for each pressure value, this leakage also exists in the \BN Raman spectra. The \BN Raman response of Hg-1201 (cf. Fig.~\ref{fig:5} (e)), being almost flat, here, it is mostly dominated by the leakage of the $A_{1g}$ contribution and of an extrinsic background common to the \BN and \BAN geometry. Using the \BN Raman response as a reference spectrum, we can subtract these contributions from the \BAN one for each pressure (for more details, see Appendix C). The pressure evolution of the \BAN electronic Raman response free of leakage is displayed in Fig.~\ref{fig:12}. It is clear that the \BAN superconducting peak energy decreases by a factor of 3 with pressure from 600 \cm to 200 \cm  (cf. Fig.~\ref{fig:12} (a)and (b)). We can notice that the evolution of the \BAN SC peak does not seem to be  continuous with pressure and it could be a pressure-plateau in between 1 and 2 GPa. 
%%%%%%%%%%%%%%%%%%%%%%%%%%%%%%%%%%%%%%%%%%%%%%%%%%%%%%%%%%%%%%%%%%%%%%%%%%%%%%%%%%%%%%%%%%%%%%%%%%%%%%%%%%%%%%%%
%%%%%%%%%%%%%%%%%%%%%%%%%%%%%%%%%%%%%%%%%%%%%%%%%%%%%%%%%%%%%%%%%%%%%%%%%%%%%%%%%%%%%%%%%%%%%%%%%%%%%%%%%%%%%%%%
The evolution with pressure of the characteristic SC peaks of the UD132 Hg-1223 and UD92 Hg-1201 are summarized in Fig.~\ref{fig:13}. In panel (a), the \BAN SC peak associated with the Hg-1223 structure has two components detected by Raman measurements at $\approx$ 0 GPa (outside cell). The two components of the SC peak (inside the cell) appear above \SI{3}{\giga\pascal}. Above this pressure, the \BAN inner plane component frequency decreases more slowly than the outer plane component. The frequency of the inner plane component does not decrease below \SI{710}{\per\centi\meter} while the one of the outer plane component continues to decrease down to \SI{590}{\per\centi\meter} at \SI{10}{\giga\pascal}. We interpret these two distinct evolutions as a different efficiency of the charge transfer with pressure between the inner and the outer planes of Hg-1223. This energy decrease is much weaker than the one of the \BAN peak in the Hg-1201 structure (cf. panel (b)). It is likely due to a larger inertia of the charge transfer induced by pressure in the triple layers than in the single one. Still, as the pressure increases, \Tc increases for both the Hg-1201 and Hg-1223 structures. The increasing of \Tc with pressure is confirmed by Raman measurements at high pressure (cf. Appendix D) that show the \BAN SC peak is still resolved well above the \Tc value measured at ambient pressure. Consequently, it is  clear that the \BAN SC peaks of Hg-1223 and Hg-1201 do not follow \Tc with pressure.  The \BN SC peak frequency detected on the  Hg-1223 spectra slightly increases with pressure following \Tc (cf. panel (a) of Fig.~\ref{fig:13}) but the $A_{1g}$ SC peak observed on the Hg-1201 spectra decreases with pressure (cf. panel (b) of Fig.~\ref{fig:13}). The $A_{1g}$ Raman data are displayed in Appendix E.\\
%%%%%%%%%%%%%%%%%%%%%%%%%%%%%%%%%%%%%%%%%%%%%%%%%%%%%%%%%%%%%%%%%%%%%%%%%%%%%%%%%%%%%%%%%%%%%%%%%%%%%%%%%%%%%%%%
%%%%%%%%%%%%%%%%%%%%%%%%%%%%%%%%%%%%%%%%%%%%%%%%%%%%%%%%%%%%%%%%%%%%%%%%%%%%%%%%%%%%%%%%%%%%%%%%%%%%%%%%%%%%%%%%
To summarize this part, the most important results are that (i) the SC \BAN peak for both  Hg-1201 and Hg-1223 decreases drastically in frequency while \Tc increases with pressure, (ii)  In the case of Hg-1201 single layer (whose charge transfer is not altered by multi-layers), the \BAN SC peak energy collapses from 10 to 2 $K_B\Tc$ (cf. panel (c) of Fig.~\ref{fig:13}). The two components of the \BAN SC peak of Hg-1223 also decreases in frequency but more slowly from 10 and 6 $K_B\Tc$. These frequency changes are significantly larger than those obtained over the same pressure range for the \BAN SC peak of YBa$_2$Cu$_3$O$_{7-\delta}$ (Y-123) \cite{Goncharov2003}. This is due to the fact that mercury-based compounds are highly compressible along the $c$-axis compared to the other cuprates\cite{Eggert1994}. It is therefore apparent that the binding energy of the Cooper pairs at its maximum value, corresponding to the \BAN pair breaking peak, does not scale with \Tc under pressure. It has already been shown than the \BAN peak energy scale detected in the Raman spectra decreases with doping like the pseudogap energy scale in several cuprate families \cite{Opel2000,Tallon01,Bernhard2008, Munnikes2011,Loret2020}, suggesting that the \BAN peak energy scale and the pseudo-gap energy scale could be linked at least as a function of doping. Can such a link be made as a function of pressure ? The hypothesis was already considered in an earlier Raman study under pressure on Y-123 \cite{Goncharov2003} although at the time it had not been established that the SC \BAN  peak did not follow \Tc with pressure as we report here. The authors had based themselves on the assumption that the pseudo-gap would be related to magnetic correlations \cite{Scalapino2012,Scalapino1995,Lee06,Anderson2007,Chubukov2008,Norman2011,Keimer2015} which are weakened as the hole concentration increase with pressure and thus, the pseudogap energy scale should decrease. Based on this hypothesis, the \BAN peak softening with pressure was interpreted as a sign of its connection to the pseudogap. This scenario deserves to be explored. Unfortunately, there are very few pressure dependent studies of the pseudogap  in the literature and the results are contradicting. Some data advocate in favor of \Ts as independent of pressure \cite{Doiron-Leyraud2017,Cyr-Choiniere2018}, while others that \Ts increases \cite{Mello2002,Vovk2018}, or decreases with pressure \cite{Hafliger2006}. Additionally, to our knowledge, no direct measurement of the pseudogap energy scale with pressure has been yet carried out so far. So, it appears that no definite link between \BAN SC gap at the anti-nodes and the pseudogap energy scale can yet be made as a function of pressure. Therefore, it would be interesting to investigate the pseudogap energy scale with pressure in order to clarify its relationship with the binding energy of the SC gap at the anti-nodes. 

\section{V. Conclusion} In summary, we have performed Raman measurements under hydrostatic pressure on the Hg-1223 and Hg-1201 cuprate superconductors. Our analysis reveals that the \Tc increase with pressure is slowed down in the Hg-1223 multi-layers compared to the Hg-1201 single layer due to the inhomogeneous increase of the carrier concentration inside the three $\mathrm{CuO}_2$ layers of the Hg-1223. We find that the frequency dependence under pressure of the apical mode from which the charge transfer operates, is the same for both the Hg-1223 and Hg-1201 cuprates and controlled by the inter-plane compressibility. Last but not least, we show that the binding energy of the Cooper pairs related to the maximum amplitude of the $d-$ wave SC gap at the anti-nodes (the \BAN SC peak) decreases drastically while \Tc increases with pressure. In particular for Hg-1201, its energy collapses from 10 to 2 $K_B\Tc$, intriguingly reaching values below the weak-coupling BCS limit\cite{Bardeen57}. These new experimental facts added to the former one that the binding energy of the Cooper pairs at the anti-nodes also decreases as \Tc increases with doping, demonstrates that the binding energy of the Cooper pairs at the anti-nodes does not follow \Tc both with doping and pressure. It could be linked to the pseudogap energy scale which follows the same trend with doping \cite{Bernhard2008,Loret2020}. However, a formal proof of this conjecture requires a measurement of the pseudogap energy scale as a function of pressure.

\section{Acknowledgments} We are grateful to V. Brouet, I. Paul, M. Civelli,  A. Georges, G. Ghiringhelli, M. Greven, M. Grilli, A. Chubukov, Marc-Henri Julien, C. P\'epin, C. Proust, Y. Sidis, L. Taillefer and A.M. Tremblay for useful discussions. We thank the University of Paris, the Coll\`ege de France and the Canadian Institute for Advanced Research (CIFAR) for their support.  We acknowledge support from  the french national agency of research,  grant NEPTUN ANR-19-CE30-0019-01. 
Correspondence and request for materials should be addressed to A.S. (alain.sacuto@univ-paris-diderot.fr).

\appendix
\section{Raw Raman data of Hg-1223 and Hg-1201 under pressure} 
\begin{figure}[ht!]
\includegraphics[scale=0.4]{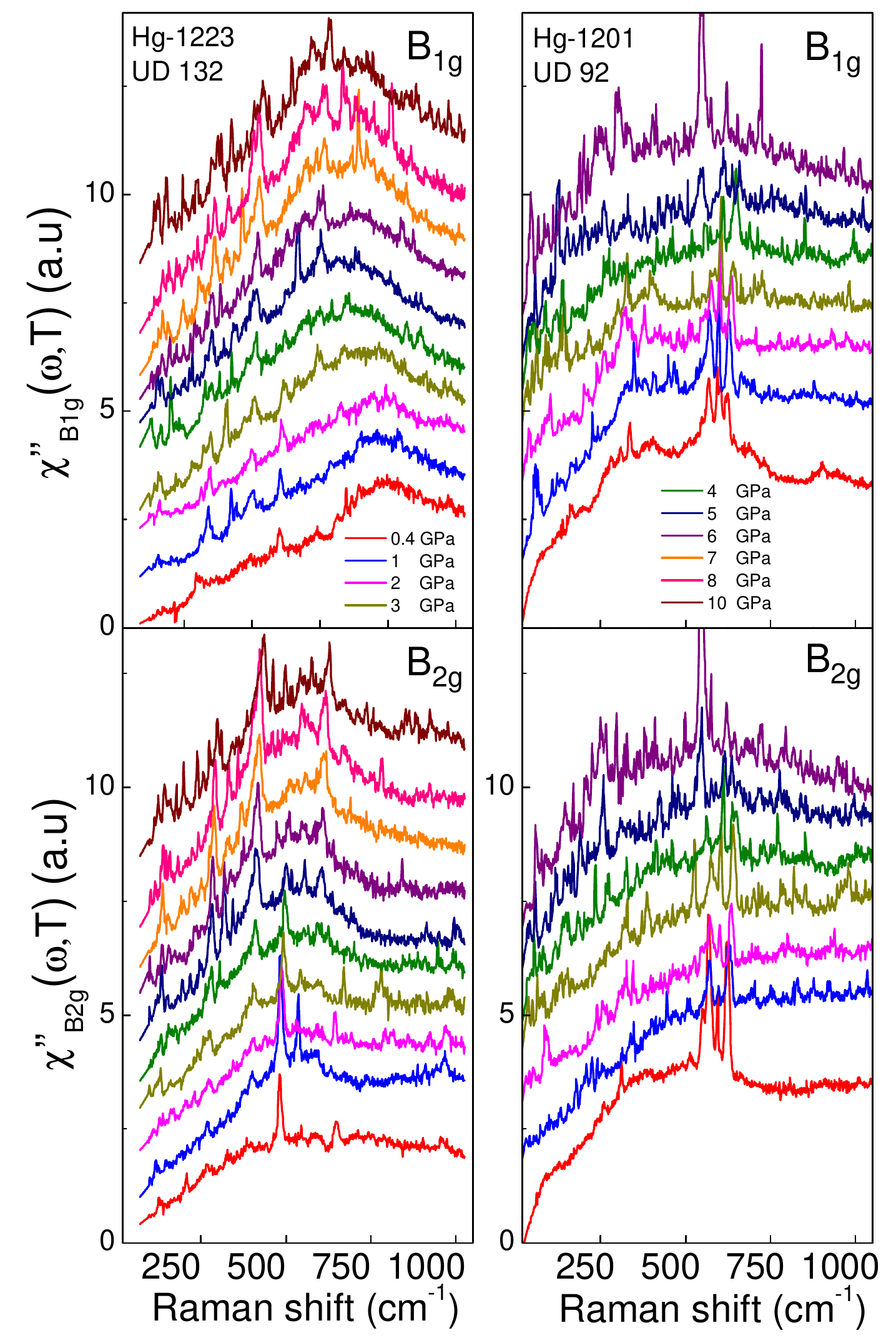}
\caption{Sets of the Hg-1201 and Hg-1223 Raman responses under hydrostatic pressures. The sets of Hg-1223 and Hg-1201 spectra were obtained at $\SI{4}{\kelvin}$ and $\SI{14}{\kelvin}$ respectively.}
\label{fig:14}
\end{figure}

The Raman superconducting response of the slightly under-doped (UD92K) Hg-1201 and (UD132K) Hg-1223 single crystals under hydrostatic pressure in \BAN and \BN geometries are reported in Fig.~\ref{fig:14}. Similarly to Fig.~\ref{fig:5} we observe a broad electronic background superimposed by few weak narrow phonon peaks stemming from pristine, parasitic and defect modes.  

%%%%%%%%%%%%%%%%%%%%%%%%%%%%%%%%%%%%%%%%%%%%%%%%%%%%%%%%%%%%%%%%%%%%%%%%%%%%%%%%%%%%%%%%%%%%%%%%%%%%%%%%%%%%%%%%
%%%%%%%%%%%%%%%%%%%%%%%%%%%%%%%%%%%%%%%%%%%%%%%%%%%%%%%%%%%%%%%%%%%%%%%%%%%%%%%%%%%%%%%%%%%%%%%%%%%%%%%%%%%%%%%%

\section{\texorpdfstring{$B_{1g}$}~ splitting under pressure in Hg-1223 } 
 
\begin{figure*}[ht!]
\includegraphics[scale=0.5]{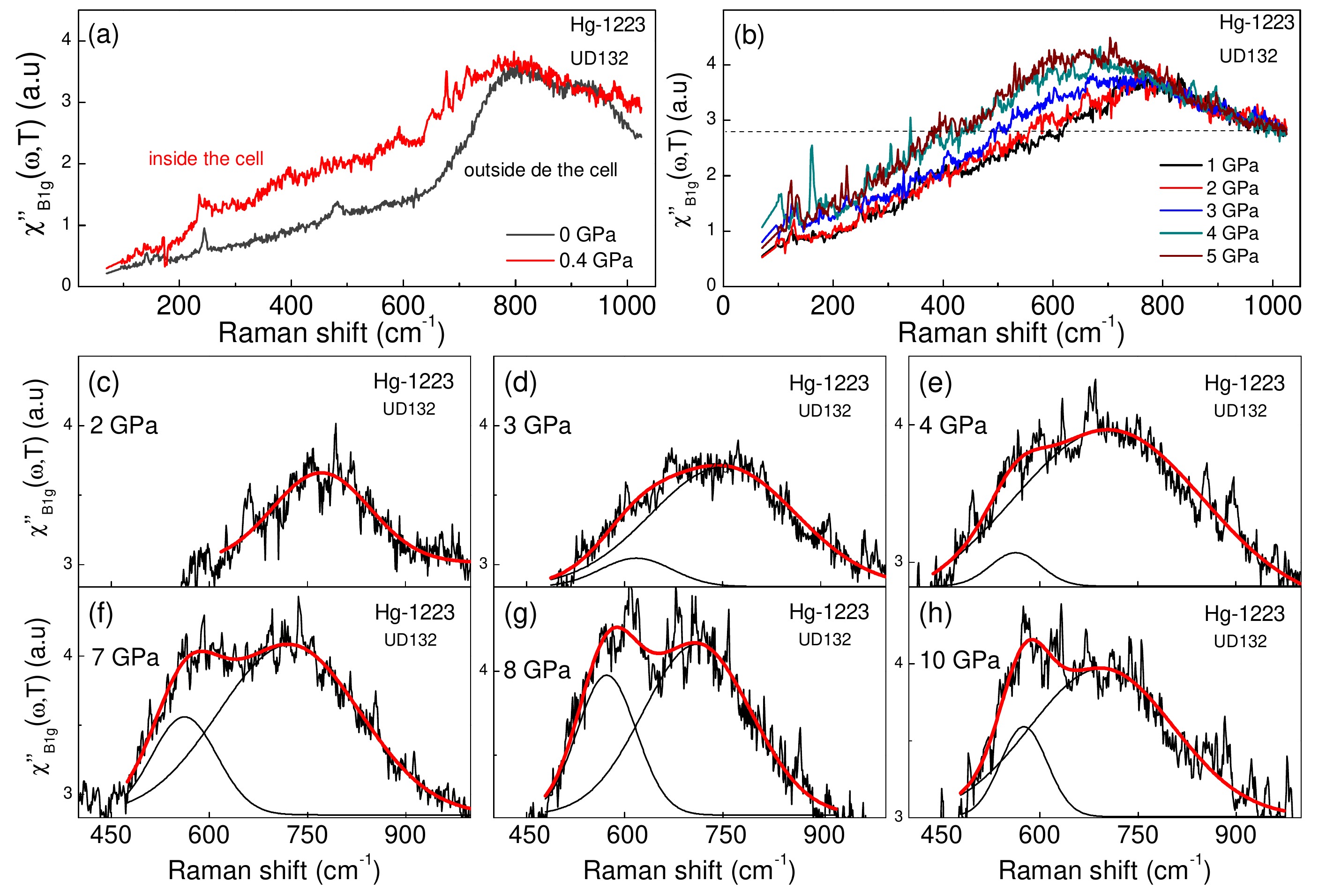}
\caption{Splitting of the \BAN SC peak under hydrostatic pressures in Hg-1223. (a) comparison between the Raman responses at 0 and 0.4 GPa measured outside and inside the pressure cell; (b) superposition  of several Raman responses of H-1223 under pressure. The based dotted line delimits the top of \BAN SC peak; (c)-(h) deconvolution of the \BAN SC peak under various pressures.}
\label{fig:15}
\end{figure*}

We have superimposed in Fig.~\ref{fig:15}, the Raman spectrum measured outside the cell at 0 GPa and the one measured inside the cell at 0.4 GPa. See panel (a). We clearly see a lower signal to noise ratio and an additional background to the Raman spectrum inside the cell. These two effects hide the weaker high energy shoulder detected on the Raman spectrum measured outside the cell and prevent us to detect the splitting at low pressure up to 2 GPa. However, as the pressure increases from 1 to 5 GPa, the linewidth of the \BAN SC peak (defined by the intersections of the left and right sides of the \BAN SC peak with the base dotted line drawn in panel (b)) increases by a factor 1.5 with pressure (cf. panel(b)). This peak broadening is likely a precursor of the observation of the splitting detected by the eyes from 6 GPa. The splitting can be revealed by a deconvolution of the \BAN SC peak by two Gaussian peaks. Below 2 GPa included, the splitting is not detected inside the cell as mentioned above. On the other hand, above 2 GPa, the fitting of the \BAN SC peak shape can only be achieved by considering two Gaussian peaks as it is shown in panels (c) to (h) where the top of the peak (above the based dotted line) for several pressures is reported. This allows us to confirm that the splitting is detectable from 3 GPa in our spectra.

%%%%%%%%%%%%%%%%%%%%%%%%%%%%%%%%%%%%%%%%%%%%%%%%%%%%%%%%%%%%%%%%%%%%%%%%%%%%%%%%%%%%%%%%%%%%%%%%%%%%%%%%%%%%%%%%
%%%%%%%%%%%%%%%%%%%%%%%%%%%%%%%%%%%%%%%%%%%%%%%%%%%%%%%%%%%%%%%%%%%%%%%%%%%%%%%%%%%%%%%%%%%%%%%%%%%%%%%%%%%%%%%%
 
\section{\texorpdfstring{$B_{1g}$}~ Raman signal under pressure in Hg-1201}
\begin{figure}[ht!]
\includegraphics[scale=0.4]{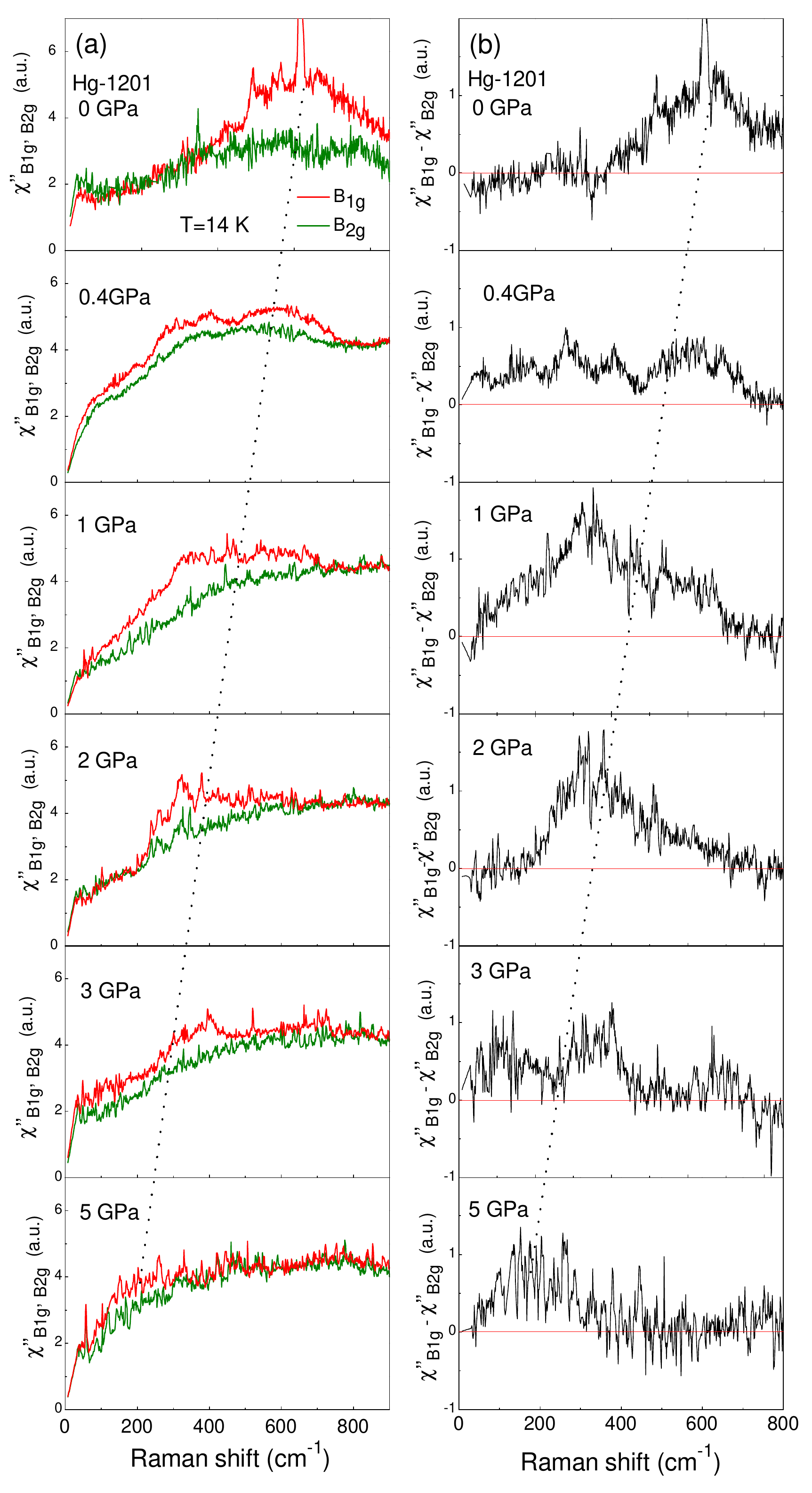}
\caption{(Color online) \BN and \BAN Raman spectra of (UD92K) Hg-1201 compounds measured as a function of pressure.}
\label{fig:16}
\end{figure}

We display in Fig.~\ref{fig:16}, panel (a), both the \BN and \BAN spectra of the (UD92K) Hg-1201 as a function of pressure measured in the superconducting state at 14 K. The subtraction of the \BN spectra from the \BAN one (cf. panel (b)), allows us to eliminate the $A_{1g}$ contribution associated with the polarization leakage induced by pressure. We see that the \BAN SC peak decreases rapidly with pressure. The 0 GPa spectra has been obtained outside the anvil cell and do not present any polarization leakage. Note that the \BN spectrum at 0 GPa does not exhibit any feature is almost flat as expected for a slightly under-doped Hg-1201 compound close to the optimal doping level as mentioned previously. 

%%%%%%%%%%%%%%%%%%%%%%%%%%%%%%%%%%%%%%%%%%%%%%%%%%%%%%%%%%%%%%%%%%%%%%%%%%%%%%%%%%%%%%%%%%%%%%%%%%%%%%%%%%%%%%%%
%%%%%%%%%%%%%%%%%%%%%%%%%%%%%%%%%%%%%%%%%%%%%%%%%%%%%%%%%%%%%%%%%%%%%%%%%%%%%%%%%%%%%%%%%%%%%%%%%%%%%%%%%%%%%%%%

\section{Raman experimental evidence of \texorpdfstring{$T_c$}~ increase with pressure in Hg-1201 and Hg-1223}

\begin{figure}[ht!]
\includegraphics[scale=0.4]{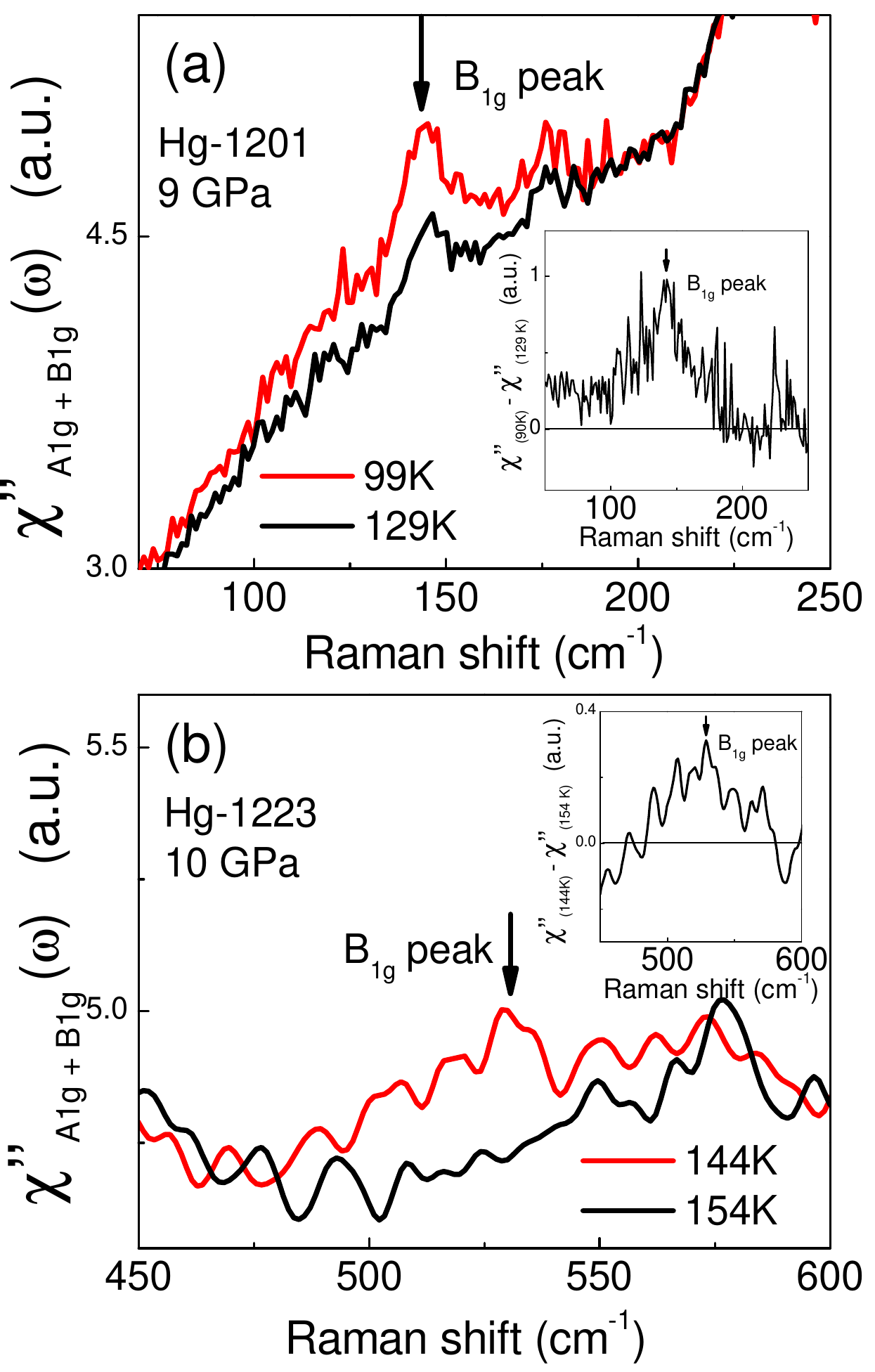}
\caption{(Color online) $A_{1g}+B_{1g}$ Raman response at high pressure for (a) (UD92K) Hg-1201 and (b) (UD132K) Hg-1223 compound. The arrows indicate the location of the \BAN SC peak. In the insets is displayed the subtraction between the Raman response above and below \Tc under pressure. The \BAN SC peak for the both compounds is still present well above \Tc at ambient pressure.}
\label{fig:17}
\end{figure}

In Fig.~\ref{fig:17} is reported the SC \BAN peak detected at 9 GPa and 10 GPa in Hg-1201 and Hg-1223 respectively. It is still observed above \Tc measured at ambient pressure (0 GPa) i.e: 92 K for Hg-1201 and 132 K for Hg-1223. Indeed, it is detected at 99 K and 144 K for Hg-1201 and Hg-1223 respectively. This means that \Tc of Hg-1201 at 9 GPa is at least greater than 99 K and that of Hg-1223 is greater than 144 K. The change in \Tc as a function of pressure for theses two compounds is then close to that found by transport measurements: namely 1 K/GPa \cite{Chu1993,nunez1993,Gao1994,Antipov2002}.
%%%%%%%%%%%%%%%%%%%%%%%%%%%%%%%%%%%%%%%%%%%%%%%%%%%%%%%%%%%%%%%%%%%%%%%%%%%%%%%%%%%%%%%%%%%%%%%%%%%%%%%%%%%%%%%%%%%%%%%%%%%%%%%%%%%%%%%%%%

\section{Evolution of the \texorpdfstring{$A_{1g}$}~  SC peak with pressure in Hg-1201}

\begin{figure}[ht]
\includegraphics[scale=0.35]{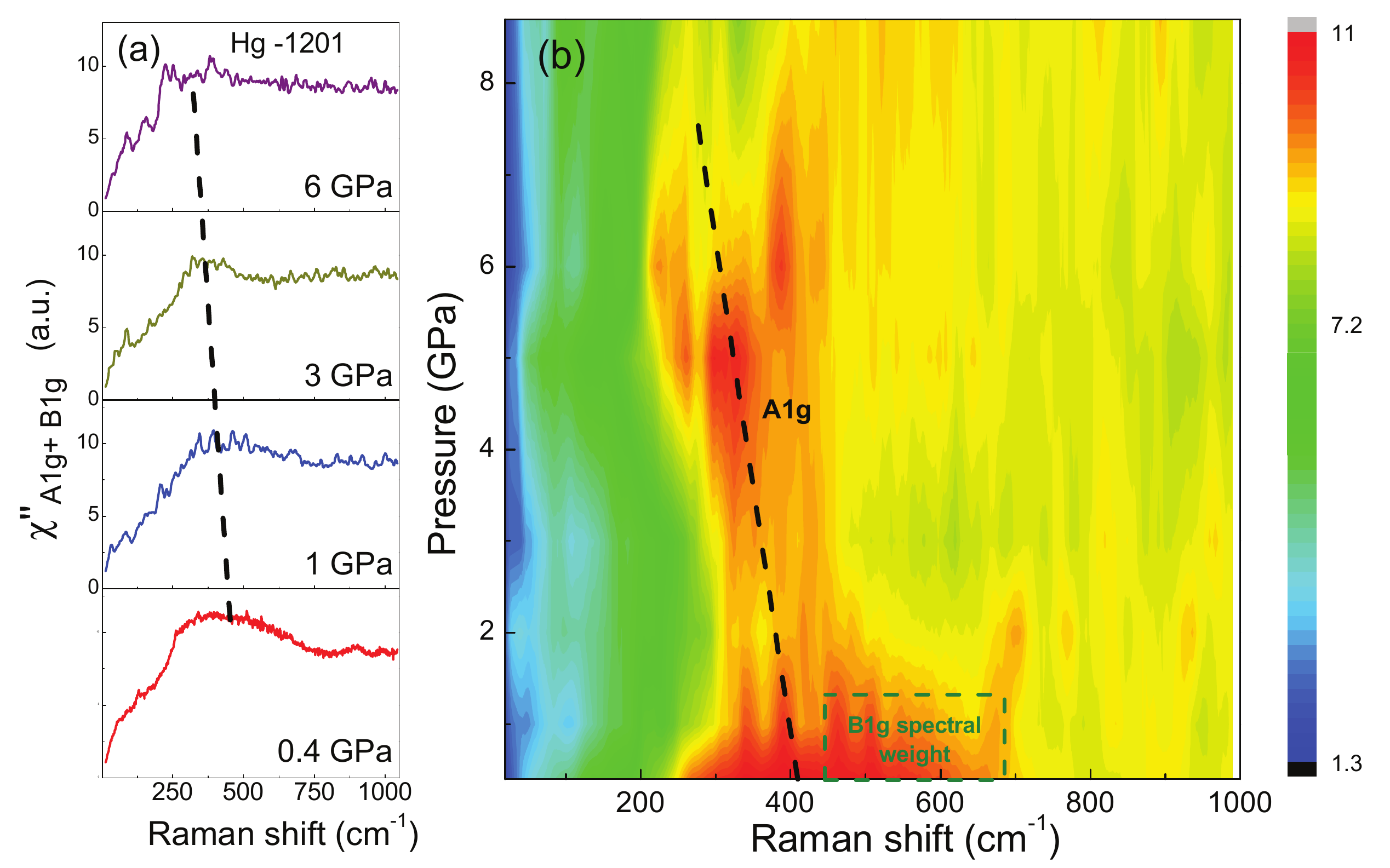}
\caption{(Color online) $A_{1g}+B_{1g}$ Raman response in Hg-1201 under pressure at $T = \SI{14}{\kelvin}$. (a)  Electronic Raman response featuring the $A_{1g}$ peak, whose frequency decreases under pressure; (b) Contour plot highlighting the evolution of the $A_{1g}$ spectral weight as a function of pressure. The dashed line is a guide for the eyes.}
\label{fig:18}
\end{figure}

We display in Fig.~\ref{fig:18}, the $A_{1g}+B_{1g}$ Raman response of the (UD92K) Hg-1201 as a function of pressure. In this geometry, the $B_{1g}$ SC peak is weak in intensity, this allows us to follow the $A_{1g}$ SC peak which decreases in energy with pressure. 

\clearpage

%* electronic address: alain.sacuto@univ-paris-diderot.fr
\bibliography{cuprates}

\end{document}